\documentclass[twocolumn,aps,prd,showpacs,nofootinbib,floatfix,eqsecnum]{revtex4}
\usepackage{amsmath,latexsym,graphicx,multirow}

\begin{document}
\title{Measuring parameters of massive black hole binaries with
  partially aligned spins} \author{Ryan N.\ Lang,$^1$ Scott
  A.\ Hughes,$^2$ and Neil J.\ Cornish$^3$}
\affiliation{$^1$Gravitational Astrophysics Laboratory, NASA Goddard
  Space Flight Center, 8800 Greenbelt Road, Greenbelt, Maryland
  20771, USA\\
  $^2$Department of Physics and MIT Kavli Institute, MIT, 77
  Massachusetts Avenue, Cambridge, Massachusetts 02139, USA\\
  $^3$Department of Physics, Montana State University, Bozeman, Montana 59717}

\begin{abstract}
The future space-based gravitational wave detector LISA will be able
to measure parameters of coalescing massive black hole binaries, often
to extremely high accuracy.  Previous work has demonstrated that the
black hole spins can have a strong impact on the accuracy of parameter
measurement.  Relativistic spin-induced precession modulates the
waveform in a manner which can break degeneracies between parameters,
in principle significantly improving how well they are measured.
Recent studies have indicated, however, that spin precession may be
weak for an important subset of astrophysical binary black holes:
those in which the spins are aligned due to interactions with gas.  In
this paper, we examine how well a binary's parameters can be measured
when its spins are partially aligned and compare results using
waveforms that include higher post-Newtonian harmonics to those that
are truncated at leading quadrupole order.  We find that the weakened
precession can substantially degrade parameter estimation.  This degradation is
particularly devastating for the extrinsic parameters sky position and
distance.  Absent higher harmonics, LISA typically
localizes the sky position of a nearly aligned binary a factor of $\sim 6$
less accurately than for one in which the spin orientations
are random.  Our knowledge of a source's sky position will thus be
worst for the gas-rich systems which are most likely to produce
electromagnetic counterparts.  Fortunately, higher harmonics of the
waveform can make up for this degradation.  By including harmonics
beyond the quadrupole in our waveform model, we find that the accuracy
with which most of the binary's parameters are measured can be
substantially improved.  In some cases, parameters can be measured as well in partially aligned
binaries as they can be when the binary spins are random.

\end{abstract}
\maketitle

\section{Introduction}
\label{sec:intro}

The coalescence of massive black hole binaries is a primary source for
the future space-based gravitational wave (GW) detector
LISA.\footnote{http://lisa.nasa.gov}  LISA will be able to detect such
sources with extremely high signal-to-noise ratio (SNR) $\gtrsim 1000$
at low redshift ($z \sim 1$), as well as with moderate, but still
reasonable, SNR $\sim 10$ at extremely high redshift ($z \sim 20$)
\cite{bmvcckk07}.  Estimated event rates for these sources vary widely
based on formation scenarios but tend to predict roughly tens of
sources per year, with $\sim 1$ as a pessimistic estimate and $\sim
100$ as an optimistic one \cite{svh07}.  (The actual detection rate
will, of course, tell us much about the formation of black hole
binaries and the growth of massive black holes in the universe.)

While the detection of gravitational waves from these sources will
certainly be interesting for its own sake, attention has turned in
recent years to the capabilities of LISA as a true astronomical
observatory.  Many papers \cite{c98, h02, v04, bbw05, hh05, bcw06, lh06,
  aiss07, aissv07, ts08, lh08, pc08, lh09, kjs09, mtbk10, kc11, mlbt11} have
investigated just how well LISA can measure the parameters of the
binaries it detects.  This is often done using the Fisher-matrix
method \cite{f92,cf94}, which essentially measures the local curvature
of the posterior probability distribution for parameters in the
vicinity of the maximum.  Parameters for which the posterior is more
strongly curved (i.e., which more strongly affect the waveform) are
measured more accurately than those for which the posterior is only
weakly curved.  Correlations between parameters are also extremely
important.  When two parameters are strongly correlated, it is
difficult to ``detangle'' the influence of one on the waveform over
the other.  This means that the accuracy with which both parameters
are measured is controlled by the one which is most poorly determined.

Recent studies have considered how well parameters can be measured
while doing the actual data analysis problem of removing signals from
noise \cite{a06}.  In effect, these studies simulate the parameter
extraction process with enough detail to uncover issues such as
multiple extrema of the posterior surfaces which are missed by the
simpler (and cruder) Fisher analyses.  Such ``realistic'' studies are
typically much more CPU-intensive and cannot easily study parameter
measurement issues over a broad swath of astrophysically important
parameter space.  Both families of studies have substantially advanced
our understanding of LISA's science reach in the past 5 or so years.

Some parameters that are especially interesting are the {\it
  intrinsic} system properties, namely, the masses and spins of the
black holes.  Masses can be measured extremely well in the best cases
[with a relative error of $\sim 10^{-3}$ for individual masses and
  $\sim 10^{-5}$ for the chirp mass $\mathcal{M} \equiv
  m_1^{3/5}m_2^{3/5}/(m_1+m_2)^{1/5}$].  Spins are not measured quite
so well but are still expected to be determined with percent-level
accuracy.  By measuring these parameters for many systems, one can
construct a merger history of black holes, and by extension, their
host galaxies, learning much about galaxy formation, black hole
formation, AGN feedback, and so forth.

We are also interested in measuring parameters {\em extrinsic} to the
system, namely, its position on the sky and its luminosity distance.
With the position and distance (converted into an approximate
redshift), astronomers can search the sky for probable electromagnetic
counterparts to the gravitational wave events.  Various types of
counterparts have been proposed, from signals during the inspiral
\cite{an02}, to bright flashes at the time of merger \cite{bp09} (or
even reductions in luminosity \cite{ombrs09}), to long-delayed
afterglows \cite{mp05}. The different scenarios arise because the
behavior of gas around an inspiraling binary system is not well
understood.  A very different kind of electromagnetic counterpart can
be produced by a kicked remnant black hole that triggers a telltale
sequence of stellar disruptions \cite{sl11}.  (Tidal disruption of stars may also allow us to flag the presence of a binary long before it enters the LISA band, allowing a better understanding of the space density of massive black hole binaries \cite{wb10}.)  If a counterpart
can be identified, the electromagnetic information can be combined
with the gravitational information to reveal more about the
astrophysics of the system.  Counterparts may also make it possible
for binary black holes to be used as probes of the cosmological
distance-redshift relation, since the electromagnetic redshift and
gravitational distance are determined independently \cite{hh05}.
Unfortunately, finding a counterpart, even if a unique signature does
exist, will not be easy.  The typical error windows for LISA are
$\sim$ tens of arcminutes on a side at the end of inspiral, reduced
from several square degrees in the weeks and months before merger.
Still, this localization does give large survey telescopes like LSST
(field of view $\sim 10$ square degrees) a chance to study a
particular area of the sky with advance warning \cite{t02}.

One particularly important result in the study of LISA's science
capabilities was the discovery that including spin precession effects
in the waveform model typically improves the accuracy of parameter
measurement \cite{v04,lh06}.  Spin precession arises because of
geodetic and gravitomagnetic general relativistic effects
\cite{acst94,k95}.  The orbital plane of the system also precesses in
order to preserve the total angular momentum on time scales shorter
than the radiation reaction time.  Together, these precessions
modulate the amplitude and phase of the waveform, breaking
correlations between certain sets of parameters and improving how well
the members of those sets can be measured.  The greatest improvement
is to the measured masses of a binary's members (accuracy typically
improved by 1--2 orders of magnitude).  The measured sky position
angles and distance to the source are all improved by about half an
order of magnitude, reducing the size of the sky position pixel in
which one must search for a counterpart by a factor of $\sim 10$ (or
the 3D voxel volume by a factor of $\sim 30$).

The precession of one of the spins in a binary, to 1.5 post-Newtonian order and averaged over an orbit,\footnote{The validity of averaging over an orbit can be quickly checked by comparing the precession time scale $T_{\rm prec}$ with the orbital time scale $T_{\rm orb}$.  For an equal-mass system, these time scales are roughly equivalent when $r \sim 7M/8$, where $M$ is total mass.  The orbit-averaged approximation is therefore quite good for most of the inspiral waveform but will begin to break down as we approach $r \sim (2-3) M$.  At these radii, the post-Newtonian approximation breaks down as well, so our use of the precession equations there is best considered to be qualitatively indicative of the relevant physics, if not numerically exact.} is given
by
\begin{equation}
  \begin{split}
    \mathbf{\dot{S}}_1 &=
    \frac{1}{r^3}\left[\left(2+\frac{3}{2}\frac{m_2}{m_1}\right)\mu
    \sqrt{Mr} \mathbf{\hat{L}}\right] \times \mathbf{S}_1 \\ & \quad +
    \frac{1}{r^3}\left[\frac{1}{2}\mathbf{S}_2-\frac{3}{2}(\mathbf{S}_2\cdot
    \mathbf{\hat{L}})\mathbf{\hat{L}}\right] \times \mathbf{S}_1 \, ,
  \end{split}
  \label{eq:S1dot}
\end{equation}
where $\mathbf{S}_1$ and $\mathbf{S}_2$ are the two spins, $m_1$ and
$m_2$ are the two masses, $M = m_1+m_2$ is the total mass, $\mu =
m_1m_2/M$ is the reduced mass, $\mathbf{\hat{L}}$ is the direction of
the orbital angular momentum, and $r$ is the orbital separation in
harmonic coordinates.  It is clear that precession is maximal when the
spins of the system are orthogonal to the orbital angular momentum and
to each other and vanishes when the spins and orbital axis are
aligned.  In \cite{lh06} (hereafter Paper I), it was assumed that the
relative orientation of the spins and the orbital angular momentum was
completely arbitrary.  The results of that paper are summarized in a
series of histograms describing parameter measurement accuracy when
the various angular momentum vectors are allowed
to point in any direction.

Recent studies have shown, however, that accreting gas in a system may
evolve the spin in such a way that the spins are at least partially
aligned with each other and with the orbit \cite{brm07,d10}.  The
degree of alignment depends on the temperature of the gas: In ``hot
gas'' models, which have polytropic index $\gamma = 5/3$, the spins
align within $30^\circ$ of the orbital axis.  ``Cold gas'' models with
$\gamma = 7/5$ align even more thoroughly, to within $10^\circ$
\cite{d10}.

Does spin-induced precession, now constrained by initial conditions,
still break degeneracies as efficiently as described in Paper I?  Any
degradation in parameter measurement capability could have a strong
effect on the ability to find electromagnetic counterparts.  The
results of \cite{lh06,lh08,lh09} may be biased toward gas-free ``dry''
mergers, severely underestimating localization errors in gaseous
``wet'' mergers --- the very systems which we are most likely to see
electromagnetically.  The effect of alignment on mass and spin
measurements is also interesting (though arguably less so, since even
a factor of several degradation for these parameters would still imply
excellent accuracy).

The goal of this paper is to answer the question posed above.  We do
so with a Fisher-matrix analysis of parameter measurement for binaries
whose spins are partially aligned according to two wet merger models:
hot gas, which aligns the spins and orbit to within $30^\circ$, and
cold gas, which aligns to within $10^\circ$.  We demonstrate that this
degree of alignment can substantially degrade parameter accuracy but
that one can ``repair'' much of this degradation by using the ``full'' waveform
model, including harmonics beyond the leading quadrupole. In what follows, we will use the terms ``gas-free'' or ``dry''
interchangeably with the term ``random spins,'' ``hot gas''
interchangeably with the phrase ``spins aligned within $30^\circ$,'' and
``cold gas'' interchangeably with the phrase ``spins aligned within
$10^\circ$.''  We also sometimes write ``$30^\circ$ alignment'' as shorthand for ``spins aligned within $30^\circ$,'' and likewise for ``$10^\circ$ alignment.''  (The two distributions contain systems with alignments less than $30^\circ$ or $10^\circ$, although with a bias toward the upper end of the allowed range.)

The outline of the paper is as follows.  We begin in
Sec.\ {\ref{sec:code}} by describing the operation of our code,
including the production of binary black hole waveforms, the LISA
response, the noise model, and how we construct the Fisher matrix.  We
focus on changes from Paper I, leaving detailed description of the
theory to that paper.

In Sec.\ {\ref{sec:results1}}, we then present results for parameter
errors in wet mergers, examining both ``hot'' and ``cold'' models.  We
compare these results to the case of dry mergers, in which spin
orientations are chosen to be completely random with respect to each
other and to the orbital angular momentum.  It should be emphasized
that throughout Sec.\ {\ref{sec:results1}}, we consider only the
leading quadrupole piece of the gravitational waveform (the so-called
``restricted'' post-Newtonian approximation).  As expected, we find
that spin alignment largely degrades LISA's ability to measure
parameters.  As a rough rule of thumb, we find that extrinsic
parameters (sky position angles and luminosity distance) are measured
a factor of $\sim 1.5-2$ less accurately for $30^\circ$ alignment and
a factor of $\sim 2-3$ less accurately for $10^\circ$ alignment.  In
the second case, alignment eliminates most of the advantage
gained by adding precession in Paper I.  We find that the impact
upon measured masses and spins depends strongly on mass ratio, with
degradation by a factor $\sim 1-3$ at $30^\circ$ alignment, and a
factor $\sim 1-9$ at $10^\circ$ alignment.  We find a handful of
cases in which partially aligned binaries actually do {\em better}
than the randomly oriented systems.  As we describe in
Sec.\ {\ref{sec:results1}}, this is due to alignment increasing these
systems' average SNR.

To combat this degradation, we introduce another degeneracy-breaking
effect.  Much early work in LISA parameter estimation made use of the
restricted waveform model, in which only the quadrupole harmonic
of the orbital phase was included and only the leading ``Newtonian''
amplitude term was used with this harmonic (although the phase was
constructed to high post-Newtonian order).  This was done because the
quadrupole harmonic dominates signal power, while the phase is the
primary source of information about the signal.  However, it has since
been shown by several groups that including higher harmonics (and
their post-Newtonian amplitudes, making the so-called full
waveform model) also breaks degeneracies and reduces parameter errors
\cite{aiss07, aissv07, ts08, pc08}.  The magnitude of the effect is
comparable to the improvement seen by including spin precession.
Recently, Klein {\it et al.}\ have presented an analysis combining both spin
precession and higher harmonics \cite{kjs09}.  A similar analysis,
based on an earlier version of our own code, was conducted by the LISA
Science Team to investigate the science reach of the LISA
mission; the results of this study are summarized in
Ref.\ {\cite{a09}}.

In Sec.\ {\ref{sec:results2}}, we replace the leading quadrupole
waveform with the full waveform.  For the case of random spins, our
answers can be compared (with some caveats) to the results of
\cite{kjs09}.  We also compute the errors for wet, partially aligned
binaries with the full waveform.  When higher harmonics are included,
parameter errors for partially aligned binaries are often no worse, or
even better, than for the case of random spins and no higher
harmonics.  In these particular cases, higher harmonics can more than
make up for the degraded impact of spin precession.  We find this
degree of improvement for the minor axis of the sky position error
ellipse and for the luminosity distance in a majority of (mass) cases.
The improvement is not quite so
good for the major axis: Although higher harmonics can reduce errors by
factors of $\sim 2$ or more, this often does not completely make up
for the loss of precession, especially at $10^\circ$ alignment.  Errors in the measured spin behave
similarly to the major axis --- their measurement is improved, but not
enough to fully compensate for the impact of aligned spins.
By contrast, we find that higher harmonics {\em always} improve mass
measurements beyond what can be done with random spins alone.  In
fact, partial alignment in many cases {\it improves} mass
measurements, thanks to increased SNR in these cases.

We also briefly take a more detailed look at the relative improvement
from spin precession, higher harmonics, and their combination.  We
confirm previous results that, for extrinsic parameters, the impact of
the combined effects is not substantially greater than the impact of
each effect alone.  For mass errors, the higher harmonics dominate,
with precession being almost irrelevant for the full waveform.  For
spin errors, however, the two effects do seem to be independent, with
the combined improvement approximately equal to (or greater than) a
simple multiplication of the individual improvements.

We conclude in Sec.\ {\ref{sec:conclusions}} by summarizing our
results and discussing additional studies that must be done before the
question of LISA parameter estimation is fully understood.  Throughout
this paper, we use geometrized units in which $G = c = 1$.  A useful
conversion factor is that $1\ M_\odot = 4.91 \times
10^{-6}\ \mathrm{s} = 1.47\ \mathrm{km}$.

Since this paper was originally written, budget constraints have caused a rescoping of the LISA mission, and the
mission that eventually flies may differ from the ``classic'' configuration considered here. We continue to focus our analysis
on measurements using LISA Classic for two reasons.  First, the design of the rescoped mission is in flux.  Until a design is fixed and its associated sensitivity known, we cannot study how well it will make measurements.  Second, our goal is to make comparisons with previous studies that were based on the classic design.  As such, it is most appropriate for us to use this design as well.  We note that our conclusions should be robust in the sense that the general trends we find regarding the impact of spins and higher harmonics will be relevant to any LISA-like design (at least for designs that have five or six links, so both waveform polarizations can be simultaneously measured).

\section{Parameter estimation code}
\label{sec:code}

The code used in this paper is a version of the \textsc{montana-mit} code
used by the LISA Parameter Estimation Taskforce \cite{a09}, updated
with some new features and bug fixes.\footnote{We note that the codes
  used in \cite{a09} were all found to produce the same answers
  provided they used the same noise models, the same signal cutoffs,
  and so on.  Our code has been well-tested in as much as other codes
  had the same features to compare against.}  In this section, we
describe the relevant features of the code, especially how it differs
from the code of Paper I \cite{lh06}.  We refer the reader to Paper I
for more detailed discussion of the waveform and parameter estimation
theory.

\subsection{Massive black hole binary waveform}

The waveform from a massive black hole binary coalescence is
traditionally divided into three distinct phases: (1) the {\em
  inspiral} of the two holes, which can be described by the
post-Newtonian expansion of general relativity; (2) the {\em merger}
of the two holes into a common event horizon, describable only by full
numerical relativistic simulations; and (3) the {\em ringdown} of the
final hole into the stationary Kerr solution, which can be described
by black hole perturbation theory.  In this work, we consider only the
inspiral, which for LISA sources can last for months to years,
accumulating large amounts of SNR and parameter information.  Because
of this fact, as well as the ease of using the post-Newtonian
approximation, inspiral-only waveforms have traditionally been used in
most, though not all, LISA parameter estimation studies.  Ringdown
information was first studied on its own by Berti, Cardoso, and Will
\cite{bcw06}.  More recently, McWilliams {\it et al.}\ added both the merger
and ringdown to the inspiral, albeit for nonspinning
binaries with an {\em a priori} known mass ratio \cite{mtbk10,mlbt11}.  They
showed that the merger can add a significant amount of parameter
information, about a factor of 3 improvement in measurement accuracy
for all parameters but mass.  Work in progress will consider the
impact of an unknown mass ratio, as well as spins.

The inspiral waveform can be described by 17 parameters: the masses of
the black holes, $m_1$ and $m_2$; their dimensionless spins,
$\chi_1 = |\mathbf{S}_1|/m_1^2$ and $\chi_2 =
|\mathbf{S}_2|/m_2^2$; the spin angles at some particular reference
time $t_0$, $\cos \theta_{S_1}(t_0)$, $\cos \theta_{S_2}(t_0)$,
$\phi_{S_1}(t_0)$, and $\phi_{S_2}(t_0)$; the orientation angles of the
orbital angular momentum at $t_0$, $\cos \theta_L(t_0)$ and
$\phi_L(t_0)$; the eccentricity $e$; the periastron angle $\gamma$;
the position of the binary on the sky, $\cos \theta_N$ and $\phi_N$;
the luminosity distance $D_L$; a reference time $t_\mathrm{ref}$
(possibly different from $t_0$); and a reference phase
$\Phi_\mathrm{ref} = \Phi(t_\mathrm{ref})$.  In this work, we assume
quasicircular orbits, eliminating $e$ and $\gamma$ and reducing the
parameter set to 15.  This assumption is also quite common, since
radiation reaction has long been expected to circularize binaries
\cite{p64}.  It should be noted, however, that recent studies indicate
that gas \cite{an05,caab09} and/or stellar interactions \cite{s10} may
cause binaries to retain a small, but significant, residual
eccentricity when they enter the LISA band.  Recent work by Key and
Cornish \cite{kc11} investigates the impact of this residual
eccentricity using a nontrivial extension of our code.

In Paper I, we used the post-Newtonian parameters $t_c$ and $\phi_c$
as the reference time $t_\mathrm{ref}$ and phase $\Phi_\mathrm{ref}$.
These parameters are, respectively, the time and phase when the
post-Newtonian frequency formally diverges.  However, Paper I made a
slight error in determining the post-Newtonian frequency and phase.
To understand this error and how to correct it, begin with the time
derivative of orbital angular frequency $\omega = 2\pi f_{\rm orb}$ (shown here to second post-Newtonian order)
\begin{equation}
\begin{split}
\frac{d\omega}{dt} &= \frac{96}{5}\frac{\eta}{M^2}(M\omega)^{11/3}\left[1-\left(\frac{743}{336}+\frac{11}{4}\eta\right)(M\omega)^{2/3} \right. \\
& \quad + (4\pi-\beta)(M\omega) + \left(\frac{34103}{18144} + \frac{13661}{2016}\eta \right. \\
& \quad \left. \left. +
\frac{59}{18}\eta^2 + \sigma \right)(M\omega)^{4/3}\right] \, ,
\end{split}
\label{eq:PNdfdt}
\end{equation}
where $\eta = \mu/M$ is the reduced mass ratio, $\beta$ is a
spin-orbit coupling term, and $\sigma$ is a spin-spin coupling term.
Exact expressions for $\beta$ and $\sigma$ are given in Paper I.
Equation \eqref{eq:PNdfdt} must be integrated once to obtain $\omega(t)$
and twice for the orbital phase $\Phi_{\rm orb}(t)$.  When the spins do not precess,
this integration can be done analytically to some specified
post-Newtonian order.  In Paper I, the analytic results were used, but
with the time-dependent expressions for $\beta$ and $\sigma$ plugged
in at the end of the process.  This is technically not correct: The
time-dependent spins should be inserted into \eqref{eq:PNdfdt}, and
then that expression should be numerically integrated to produce
$\omega(t)$ and $\Phi_{\rm orb}(t)$.  This is not difficult, only requiring two
additional differential equations in the Runge-Kutta solver of Paper I.
However, it means that $t_c$ and $\Phi_c$ are no longer acceptable
references, since the numerical integrator cannot reach infinite
frequency.  We describe our current approach momentarily.

Another change from the code used in Paper I is in the choice of
cutoff frequency for the inspiral.  In Paper I, the inspiral was
stopped at the frequency of the Schwarzschild innermost stable
circular orbit (ISCO), $r = 6M$.  This assumption is poor for two
reasons.  First, while $r = 6M$ is the ISCO for a test particle
orbiting a single Schwarzschild hole of mass $M$, the dynamics of the
two-hole system are much more complex, and the transition to plunge
and merger is not so well-defined.  Second, we are considering Kerr
black holes, for which even in the point-particle limit the innermost
stable orbit can vary from $r = 9M$ to $r = M$ depending on the spin
of the hole, with a concomitantly wide variation in the ISCO
frequency.  A better solution is to stop the inspiral at the minimum
energy circular orbit (MECO), the orbit which minimizes the expression
for post-Newtonian energy \cite{bcv03}:

\begin{equation}
\begin{split}
&E = -\frac{\mu}{2}(M\omega)^{2/3}\left(1-\frac{1}{12}(9+\eta)(M\omega)^{2/3} \right. \\
& +\frac{8}{3M^2}\left[\left(1+\frac{3}{4}\frac{m_2}{m_1}\right)\mathbf{\hat{L}}\cdot \mathbf{S}_1+\left(1+\frac{3}{4}\frac{m_1}{m_2}\right)\mathbf{\hat{L}}\cdot \mathbf{S}_2\right](M\omega) \\
& +\left[\frac{1}{24}\left(-81+57\eta-\eta^2\right) \right. \\
& + \left. \left. \frac{1}{\eta M^4}(\mathbf{S}_1\cdot \mathbf{S}_2-3(\mathbf{\hat{L}}\cdot \mathbf{S}_1)(\mathbf{\hat{L}}\cdot \mathbf{S}_2))\right](M\omega)^{4/3}\right) \, .
\end{split}
\end{equation}
The MECO is known to be a better approximation to the inspiral-plunge
transition than the ISCO, and it properly takes spins into account.

Using the MECO gives us a better reference point for our time and
phase than the coalescence time and phase $t_c$ and $\Phi_c$ described
above.  We choose $t_0 = t_\mathrm{ref} = t_\mathrm{MECO}$ and 
$\Phi_\mathrm{ref} = \Phi_\mathrm{MECO}$ and then integrate the
spin, frequency, and phase evolution equations {\em backwards} from
the MECO to $t = 0$.  The backwards
integration provides stability in the Fisher-matrix calculation: We
align the waveforms when they are largest, thus making it easier to
introduce slight perturbations.

As seen in Eq.\ \eqref{eq:PNdfdt}, we calculate the phase out to
second post-Newtonian (2PN) order.  (By numerically integrating
\eqref{eq:PNdfdt} to obtain $\omega(t)$ and $\Phi_{\rm orb}(t)$, we specifically are
choosing the ``TaylorT4'' PN approximant \cite{b07}.)  We integrate
the spin precession equations out to 1.5PN order, which includes 1PN
spin-orbit and 1.5PN spin-spin terms.  It is worth noting that all of
the relevant quantities are known to higher post-Newtonian order.  Work
in preparation shows that including terms beyond the order we include
here only causes a slight quantitative change in the accuracy with
which parameters are measured {\cite{osh11}}.  In
Sec.\ {\ref{sec:results1}}, we use the restricted post-Newtonian
approximation, in which we only consider the quadrupole term ($\Phi =
2\Phi_{\rm orb}$) with its lowest order, Newtonian amplitude.  In
Sec.\ {\ref{sec:results2}}, we use the full post-Newtonian
waveform, which includes all harmonics to 2PN order in amplitude.

\subsection{LISA response and noise}

The LISA response used in this paper differs from the response used in
Paper I.  For signals which do not reach above $3 \times 10^{-3}$ Hz,
we use the same low-frequency approximation used in that paper.  In
this approximation, we ignore the transfer functions which arise due to
the finite arm lengths of the detector.  This approximation is very
inaccurate above $f \sim 3 \times 10^{-3}$ Hz.  With the addition of
higher harmonics, many signals now reach into this range where the
transfer functions become important.

The full LISA detector response is somewhat complicated to model.
Three existing codes provide the full response: the LISA Simulator
\cite{rcp04}, Synthetic LISA \cite{v05}, and LISACode
\cite{pahjpprv08}.  Interfacing with one of these codes would
significantly slow our analysis, making it difficult to perform large
Monte Carlo studies over our parameter space.  We seek a simpler
response function which includes the finite arm length transfer
functions but ignores some of the more complicated issues.

The three LISA spacecraft follow eccentric orbits around the Sun at 1
AU.  The individual orbits combine in such a way that the LISA
constellation maintains, at first order in orbital eccentricity, an
equilateral triangle formation.  By going beyond this leading order,
one finds that the arm lengths vary by a small amount on monthlong
time scales.  The variation in LISA arm lengths is the reason for the
development of time delay interferometry (TDI) techniques \cite{aet99}
to eliminate
laser phase noise, which cancels exactly in equal-arm interferometers
like LIGO.  In our detector model, we approximate the
constellation as having arm lengths that are equal at all times.  Our
model detector is a ``rigid'' equilateral triangle.

The other complexity in the full LISA response is that the spacecraft
move during the measurement, causing ``point-ahead'' effects which
must be taken into account.  We assume instead an ``adiabatic''
detector, in which for each time that we require the detector
response, the detector is considered to be motionless for that time.
The spacecraft then adiabatically move to their next position for the
next sample point.  This rigid, adiabatic approximation is known to be
equivalent to the full response up to very high frequency ($\sim 500$
mHz) and thus is appropriate for our Fisher-matrix analysis
\cite{rcp04}.

For the rigid, adiabatic approximation, the code produces Michelson
variables $X$, $Y$, and $Z$ as defined in \cite{v05}, Eqs.\ (10) and
(11).  Note that these are technically {\em not} TDI variables.  Since
we do not have to subtract phase noise, there is no need to include
another pass through the interferometer (cf.\ the ``real'' TDI
variables in Eq.\ (13) of \cite{v05}).  They do contain the same
information, though, so we may refer to them as (pseudo, equal-arm)
TDI Michelson variables in this paper.  From them, we can construct
noise-orthogonal TDI variables $A$, $E$, and $T$, defined as
\begin{align}
A &= \frac{1}{3}(2X-Y-Z) \, ,\\
E &= \frac{1}{\sqrt{3}}(Z-Y) \, , \\
T &= \frac{1}{3}(X+Y+Z) \, .
\end{align}
$A$, $E$, and $T$ are used to calculate the SNR and the Fisher matrix.
Note that, as defined in Eqs.\ (10) and (11) of \cite{v05}, these are
{\em fractional-frequency} variables.  We can convert them to {\em
  equivalent strain} by integrating the signal in the frequency domain
and then multiplying by $c/(4\pi L)$.  For the low-frequency case, we
use the Michelson signal $X$ and the noise-orthogonal signal
$(X+2Y)/\sqrt{3}$.  These combinations are denoted $h_I$ and $h_{II}$
in Paper I (which in turn follows the convention of Cutler
{\cite{c98}}).  The low-frequency approximation is constructed so that
these signals are already expressed as equivalent strain.

The LISA noise power spectral density $S_n(f)$ comprises two parts,
instrumental noise and confusion noise due to unresolved white dwarf
binaries in the Galaxy.  Instrumental noise consists of both position
noise, due to photon shot noise and other effects along the optical
path, and acceleration noise, due to proof mass motion.  The total
instrument noise in the $A$ and $E$ (strain) channels is given by
\begin{equation}
\begin{split}
S_{n,AE} &= \frac{1}{3L^2}\left[\vphantom{\frac{4S_a(f)}{2\pi f}}\left(2+\cos x\right)S_p(f) + (1+\cos x+\cos^2 x) \quad \quad
    \right.\\ &\quad  \times
    \left. \left(\frac{4S_a(f)}{(2\pi
      f)^4}\left[1+\left(\frac{10^{-4}\ \mathrm{Hz}}{f}\right)\right]\right)\right]
  \, ,
\end{split}
\label{eq:SnAE}
\end{equation}
and the $T$
(strain) noise is given by
\begin{equation}
\begin{split}
S_{n,T} &= \frac{1}{3L^2}\left[\vphantom{\frac{4S_a(f)}{2\pi f}}\left(1-\cos x\right)S_p(f) \right.\\
&\quad + \left. \frac{1}{2}\left(1-\cos x\right)^2\left(\frac{4S_a(f)}{(2\pi f)^4}\left[1+\left(\frac{10^{-4}\ \mathrm{Hz}}{f}\right)\right]\right)\right] \, .
\end{split}
\label{eq:SnT}
\end{equation}
Here $L = 5\times 10^9\ \mathrm{km}$ is the LISA arm length, $x =
2\pi fL/c$, $S_p(f) = 3.24\times 10^{-22}\ \mathrm{m}^2/\mathrm{Hz}$ is
the position noise budget, and $S_a(f) = 9\times
10^{-30}\ \mathrm{m}^2/\mathrm{s}^4/\mathrm{Hz}$ is the acceleration
noise budget.  For the
low-frequency approximation, we can calculate similar expressions for
the two noise-orthogonal channels and then take $\cos x = 1$, although
this attention to detail makes little difference for the frequencies of interest.  Notice that the position noise and acceleration noise are both assumed to be white, with no
frequency dependence.  However, because it is expected that LISA's acceleration noise performance will degrade somewhat from this white form below $10^{-4}$ Hz, we have also added a ``pink'' acceleration noise term, with a slope of $f^{-1}$.  

Confusion noise is constructed from the residuals of a fit to the
Galaxy \cite{cc07} in the Mock LISA Data Challenge \cite{a06}.  An
approximate analytic expression for the confusion noise can be found
in \cite{kc11}, Eq.\ (10).  It is added to instrument noise for the
$A$ and $E$ channels (or the orthogonal low-frequency channels) to
obtain the total noise.  It is not added to the $T$ channel because
it occurs only at low frequency, where that channel adds nothing to the
analysis.

Finally, although it is not expressed explicitly in \eqref{eq:SnAE} and \eqref{eq:SnT}, we enforce a low-frequency cutoff of $3\times 10^{-5}$ Hz and do not include any contribution
from the signal below that frequency in our analysis.  (This frequency is the lowest frequency at which LISA is planned to have good sensitivity to gravitational waves; though it will have sensitivity to sources at lower frequencies, the noise characteristics below $f = 3\times 10^{-5}$ Hz cannot be guaranteed.)

\subsection{Construction of the Fisher matrix}

The Fisher matrix $\Gamma_{ij}$ is defined as
\begin{equation}
\Gamma_{ij} = \left(\frac{\partial h}{\partial \theta^i}\left|
\frac{\partial h}{\partial \theta^j}\right.\right) \, ,
\label{eq:fishermatrix}
\end{equation}
where $h$ is the gravitational wave signal, $\theta^i$ are the 15
parameters which describe it, and
\begin{equation}
(a|b) = 4 \, \mathrm{Re} \int_0^{\infty} df
\frac{\tilde{a}^*(f)\tilde{b}(f)}{S_n(f)}
\label{eq:innerproduct}
\end{equation}
is a noise-weighted inner product.  The inverse of the Fisher matrix
is the covariance matrix, which contains squared parameter errors
along the diagonal and correlations elsewhere.  To calculate the
Fisher matrix, we need the waveforms in the frequency domain.  In
Paper I, we actually did all calculations in the frequency domain by
using the stationary phase approximation.  This approximation relies
on a separation of time scales and is known to be quite good for
nonspinning binaries, where the inspiral time scale $T_{\rm insp}$ is
much larger than the orbital time scale $T_{\rm orb}$.  However, when
precession is included in the waveform, an additional time scale
$T_{\rm prec}$ comes into play, with $T_{\rm insp} > T_{\rm prec} >
T_{\rm orb}$.  We have seen that with precession, the stationary phase
approximation tends to smooth out sharp features in the Fourier
transform, potentially reducing the information content.  The problem
becomes worse as the impact of precession increases (i.e., with higher
spin values, or for highly nonaligned spins and orbit).  To avoid
introducing any errors due to this approximation, we here calculate
our waveforms in the time domain and then perform a fast Fourier
transform (FFT) to bring them into the frequency domain.

This approach has two major limitations.  First, it is much slower
than the stationary phase approach, since we need to calculate many
time samples to observe Nyquist sampling requirements and we then need
to compute the FFT.  Second, the FFT assumes a periodic signal.
Because we have a finite signal which looks much different at the end
than at the beginning, we must introduce some kind of window in order
to taper the signal to zero at the beginning and end.  We use a Hann
window (actually half a Hann window at each end of the signal).  This
window substantially reduces ``ringing,'' or spectral leakage
problems.  However, it also cuts out part of the signal.  This is
particularly unfortunate for the strongly chirping inspiral, since
much of the signal power is contained in the last few cycles.  By
windowing the signal, we lose some of this power.  This may cause our
SNR and errors to be smaller and larger, respectively, than they would
be for a ``real,'' physical signal.
The best solution to this problem would be to include the merger and
ringdown portions of the signal, allowing it to fade to zero in a
physical, not artificial, way.  For now, we must simply accept the
windowing as part of the definition of the (unphysical) inspiral-only
waveform.

\section{Parameter estimation in partially aligned binaries: Only
the quadrupole harmonic}
\label{sec:results1}

Here we describe the parameter estimation capabilities of LISA without
including the influence of higher harmonics.  In order to consider a
wide range of LISA sources, we choose only three parameters
explicitly, the two masses of the system and the luminosity distance.
For the masses, we consider a variety of systems ranging in total mass
from $2\times 10^5\ M_\odot$ to $2\times 10^7\ M_\odot$, with a mass
ratio from $1-10$.  On the other hand, we consider only sources at $z
= 1$, corresponding to a luminosity distance of $6.64$ Gpc using our
choice of cosmological parameters.  Errors at other redshifts can be
constructed using the errors at $z = 1$.  We note that the results at
masses $m_1$ and $m_2$ and redshift $z^\prime$ can be simply related
to the results at masses $m_1(1+z^\prime)/(1+z)$ and
$m_2(1+z^\prime)/(1+z)$ and redshift $z$.  This is because all time
scales in the system are derived from the masses.  Since time scales
are lengthened (frequencies shortened) by the cosmological redshift, a
binary at higher redshift behaves like a binary at lower
redshift but with a higher mass.  The quantity $m(1+z)$ is generally
called the redshifted mass, where $m$ is the mass measured locally at
the rest frame of the binary.  (When we quote masses in this paper, we
always mean the {\em rest-frame} mass, remembering that when put into
the waveform formulas of Paper I, they must be multiplied by $1+z$.)
The amplitude of the waves at redshift $z^\prime$ is decreased by a
factor $\xi = D_L(z)/D_L(z^\prime)$ over the corresponding binary
(i.e., the binary with the same redshifted mass) at redshift $z$.
This increases the errors by $1/\xi$ over that corresponding signal.

The other 12 parameters of the system are generated essentially at
random, with 1000 different Monte Carlo realizations.  For example,
$t_{\rm MECO}$ is chosen from within an assumed three-year mission time,
meaning that some early binaries will have abnormally short signals
for a given mass.  Spin magnitudes are chosen uniformly between 0 and
$1$, and $\Phi_{\rm MECO}$ is chosen uniformly between 0 and $2\pi$.
Cosines of angles are
chosen uniformly between $-1$ and $1$, while longitudinal angles are
chosen uniformly between 0 and $2\pi$.  In the case of random spins
(as in Paper I), the procedure is then complete.  In the case of partially
aligned spins, the main focus of this paper, we use the randomly generated
parameters to integrate the 
spin precession equations backwards from the MECO to $t = 0$.  
We assume that any
alignment at $t = 0$ is solely due to gas.  If either of the resulting spin-orbit 
angles is greater than the model's restriction ($30^\circ$ 
or $10^\circ$), we randomly select new spin orientation angles (at MECO) and try again.
This procedure guarantees that all sources will have the desired amount of 
alignment at the start of the signal.  However, our sample will include some
sources ($\sim 30\%$) which move out of alignment by MECO.  Since these sources generally precess more strongly than the others, they tend to improve the overall distribution of parameter errors, especially for spin magnitude.

\begin{figure}[!b]
\includegraphics[scale=0.54]{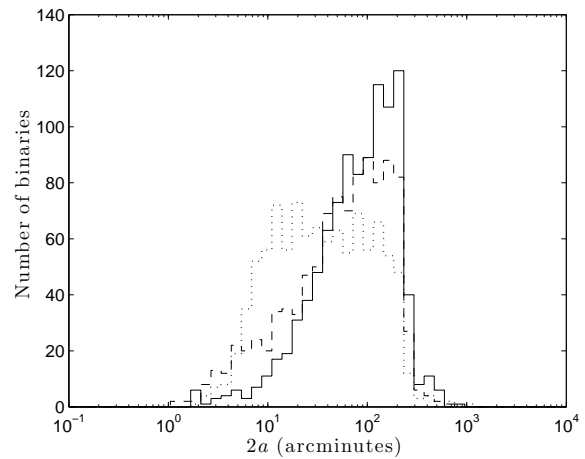}
\caption{Distribution of $2a$, the major axis of the sky position error ellipse, for
  binaries with randomly aligned spins (dotted line), spins restricted
  to within $30^\circ$ of the orbital angular momentum (dashed line),
  and spins restricted to within $10^\circ$ of the orbital angular
  momentum (solid line).  Here $m_1 = 10^6 M_\odot$, $m_2 = 3 \times
  10^5 M_\odot$, and $z = 1$.}
\label{fig:2anoHH}
\end{figure}

Figure \ref{fig:2anoHH} shows a histogram of the Monte Carlo results
for a binary with $m_1 = 10^6 M_\odot$ and $m_2 = 3 \times 10^5
M_\odot$.  We show the major axis of the sky position error ellipse,
$2a$, comparing the cases of randomly aligned spins to spins
restricted to be aligned within $30^\circ$ (for hot gas) and
$10^\circ$ (for cold gas) of the orbital angular momentum.  We see
that partial alignment of the spins and orbital angular momentum
degrades LISA's localization capability.  For the partially aligned
cases, the shapes of the histograms resemble the strongly peaked ``no
precession'' results of Paper I more than the roughly flat random-spin
histogram.  The medians of the distributions also increase: While
randomly oriented binaries have a median major axis of 34.8
arcminutes, systems aligned within $30^\circ$ have a median $2a$ of
62.3 arcminutes.  For $10^\circ$ alignment, this degrades further to
90.5 arcminutes.  This is a factor of 2.6 degradation from the
case of random alignment, just short of the factor $\sim 3$ improvement seen in Paper I
when precession is introduced into the waveform model.  In essence, by
restricting the spin angles to within $10^\circ$ of the orbital
angular momentum, we have eliminated almost all of the advantage gained from
including precession effects in the waveform.

\begin{figure}[!t]
\includegraphics[scale=0.54]{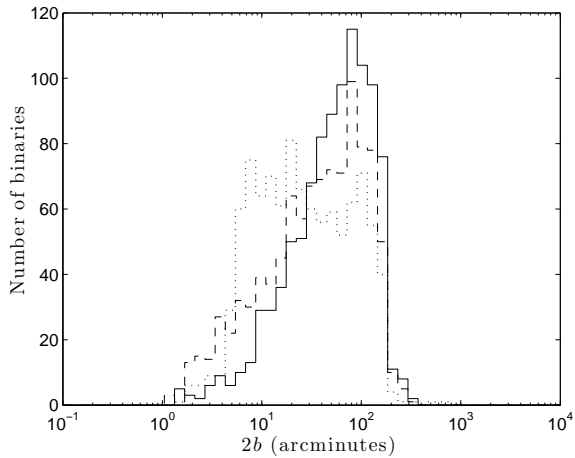}
\caption{Same as Fig. \ref{fig:2anoHH}, but for the minor axis $2b$.}
\label{fig:2bnoHH}
\end{figure}

Figure \ref{fig:2bnoHH} shows results for the minor axis of the sky
position error ellipse,\footnote{Note that unlike in Paper I, the results for $2b$ do not feature a long tail of small errors.  We have confirmed that this effect was caused by a bug in the code used in Paper I.} $2b$.  The results are similar: The median
value of $2b$ increases from 24.6 arcminutes for random spins to 40.6
arcminutes for $30^\circ$ alignment and to 58.6 arcminutes for
$10^\circ$ alignment.
Together with the results for the major axis, these numbers imply that
the total sky position area increases by a factor $> 6$
when binaries have closely aligned angular momentum vectors, strongly
impacting the ability of LISA to find electromagnetic counterparts to
the GW signal.

\begin{table}[!t]
\begin{center}
\begin{tabular}{|c|c||c|c||c|c||c|c|}
\hline
\multirow{2}{*}{$m_1\ (M_\odot)$} & \multirow{2}{*}{$m_2 (M_\odot)$} & \multicolumn{2}{|c||}{No gas} & \multicolumn{2}{|c||}{Hot gas} & \multicolumn{2}{|c|}{Cold gas} \\
& & $2a$ & $2b$ & $2a$ & $2b$ & $2a$ & $2b$ \\
\hline \hline
$10^5$ & $10^5$ & 27.0 & 16.4 & 40.7 & 25.7 & 53.8 & 35.1\\ 
\hline
$3 \times 10^5$ & $10^5$ & 17.5 & 11.7 & 30.1 & 17.9 & 53.7 & 34.8\\
\hline
$3 \times 10^5$ & $3 \times 10^5$ & 33.3 & 19.0 & 45.9 & 27.1 & 63.1 & 42.4\\
\hline
$10^6$ & $10^5$ & 23.3 & 18.3 & 35.9 & 21.6 & 61.6 & 38.4\\
\hline
$10^6$ & $3 \times 10^5$ & 34.8 & 24.6 & 62.3 & 40.6 & 90.5 & 58.6\\
\hline
$10^6$ & $10^6$ & 56.9 & 37.5 & 87.7 & 57.2 & 105 & 68.3\\
\hline
$3 \times 10^6$ & $3 \times 10^5$ & 39.0 & 33.6 & 57.0 & 36.8 & 105 & 68.1\\
\hline
$3 \times 10^6$ & $10^6$ & 45.5 & 32.4 & 83.3 & 49.0 & 131 & 77.9\\
\hline
$3 \times 10^6$ & $3 \times 10^6$ & 71.9 & 43.6 & 126 & 75.6 & 168 & 106\\
\hline
$10^7$ & $10^6$ & 47.3 & 40.2 & 70.7 & 46.6 & 132 & 83.7\\
\hline
$10^7$ & $3 \times 10^6$ & 67.3 & 45.3 & 131 & 75.7 & 234 & 143\\
\hline
$10^7$ & $10^7$ & 160 & 84.8 & 281 & 136 & 581 & 323\\
\hline 
\end{tabular}
\caption{Median sky position major axis $2a$ and minor axis $2b$, in
  arcminutes, for binaries of various masses at $z = 1$, in the ``no
  gas'' (random-spin), ``hot gas'' ($30^\circ$ alignment), and ``cold
  gas'' ($10^\circ$ alignment) cases.}
\label{table:2a2bnoHH}
\end{center}
\end{table}
Table \ref{table:2a2bnoHH} shows the major and minor sky position axes
for a range of masses, in the random-spin, hot gas, and cold gas
cases.  We see that degradation of a factor $\sim 2-3$ between the
``no gas'' and ``cold gas'' ($10^\circ$ alignment) cases occurs rather
consistently for different masses and mass ratios.

\begin{figure}[!b]
\includegraphics[scale=0.54]{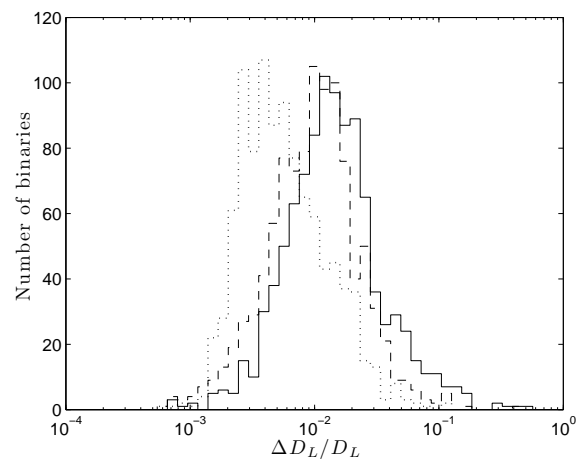}
\caption{Same as Fig. \ref{fig:2anoHH}, but for the fractional error
  in luminosity distance, $\Delta D_L/D_L$.}
\label{fig:DnoHH}
\end{figure}

\begin{table}[!t]
\begin{center}
\begin{tabular}{|c|c||c|c|c|}
\hline
$m_1\ (M_\odot)$ & $m_2 (M_\odot)$ & No gas & Hot gas & Cold gas \\
\hline \hline
$10^5$ & $10^5$ & $4.16\times 10^{-3}$ & $7.97\times 10^{-3}$ & 0.0130\\ 
\hline
$3 \times 10^5$ & $10^5$ & $2.59\times 10^{-3}$ & $6.03\times 10^{-3}$ & 0.0101\\
\hline
$3 \times 10^5$ & $3 \times 10^5$ & $5.54\times 10^{-3}$ & $9.23\times 10^{-3}$ & 0.0121\\
\hline
$10^6$ & $10^5$ & $3.67\times 10^{-3}$ & $5.92\times 10^{-3}$ & 0.0113\\
\hline
$10^6$ & $3 \times 10^5$ & $5.24\times 10^{-3}$ & 0.0101 & 0.0136\\
\hline
$10^6$ & $10^6$ & $9.37\times 10^{-3}$ & 0.0137 & 0.0175\\
\hline
$3 \times 10^6$ & $3 \times 10^5$ & $5.58\times 10^{-3}$ & $9.19\times 10^{-3}$ & 0.0147\\
\hline
$3 \times 10^6$ & $10^6$ & $7.06\times 10^{-3}$ & 0.0130 & 0.0191\\
\hline
$3 \times 10^6$ & $3 \times 10^6$ & 0.0137 & 0.0207 & 0.0279\\
\hline
$10^7$ & $10^6$ & $7.67\times 10^{-3}$ & 0.0135 & 0.0242\\
\hline
$10^7$ & $3 \times 10^6$ & 0.0129 & 0.0243 & 0.0429\\
\hline
$10^7$ & $10^7$ & 0.0441 & 0.0613 & 0.0974\\
\hline 
\end{tabular}
\caption{Same as Table \ref{table:2a2bnoHH}, but for the fractional error in luminosity distance, $\Delta D_L/D_L$.}
\label{table:DnoHH}
\end{center}
\end{table}

The other extrinsic parameter of interest is the luminosity distance
$D_L$.  Figure \ref{fig:DnoHH} shows the fractional errors in $D_L$
for different degrees of spin alignment.  Again, we see that
restricting the spin angles dramatically affects measurement: The
median of $5.24\times 10^{-3}$ for random spin orientation doubles to
$1.01\times 10^{-2}$ when the spins are aligned within $30^\circ$ and nearly
triples to $1.36\times 10^{-2}$ when the spins are aligned within
$10^\circ$.  However, this particular degradation is
almost certainly immaterial, at least at low redshift, since the error
remains much smaller than the $\sim 5\%$ error produced by weak
gravitational lensing at $z \sim 1$.  For sources at higher redshift,
this degradation may be more important.  Table \ref{table:DnoHH} shows
luminosity distance errors for different masses.  Like the sky
position, the degradation is about a factor of $\sim 1.5-2$ for
$30^\circ$ alignment and $\sim 2-3$ for $10^\circ$ alignment.  Note
that for the larger masses we consider, the degradation pushes the GW
distance error to a value comparable to or even larger than the weak
lensing error.

\begin{figure}[!b]
\includegraphics[scale=0.54]{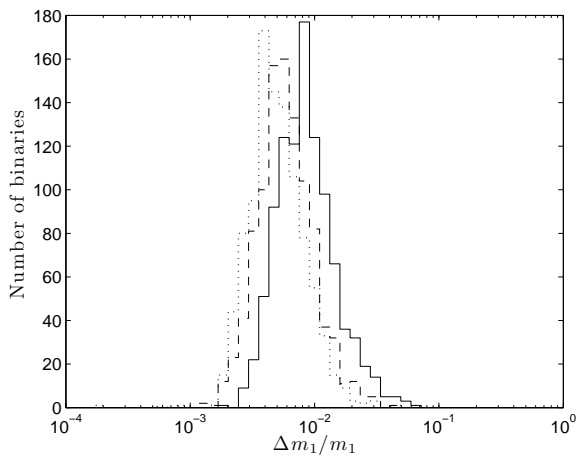}
\caption{Same as Fig. \ref{fig:2anoHH}, but for the fractional error in mass, $\Delta m_1/m_1$.}
\label{fig:m1noHH}
\end{figure}

\begin{figure}[!t]
\includegraphics[scale=0.54]{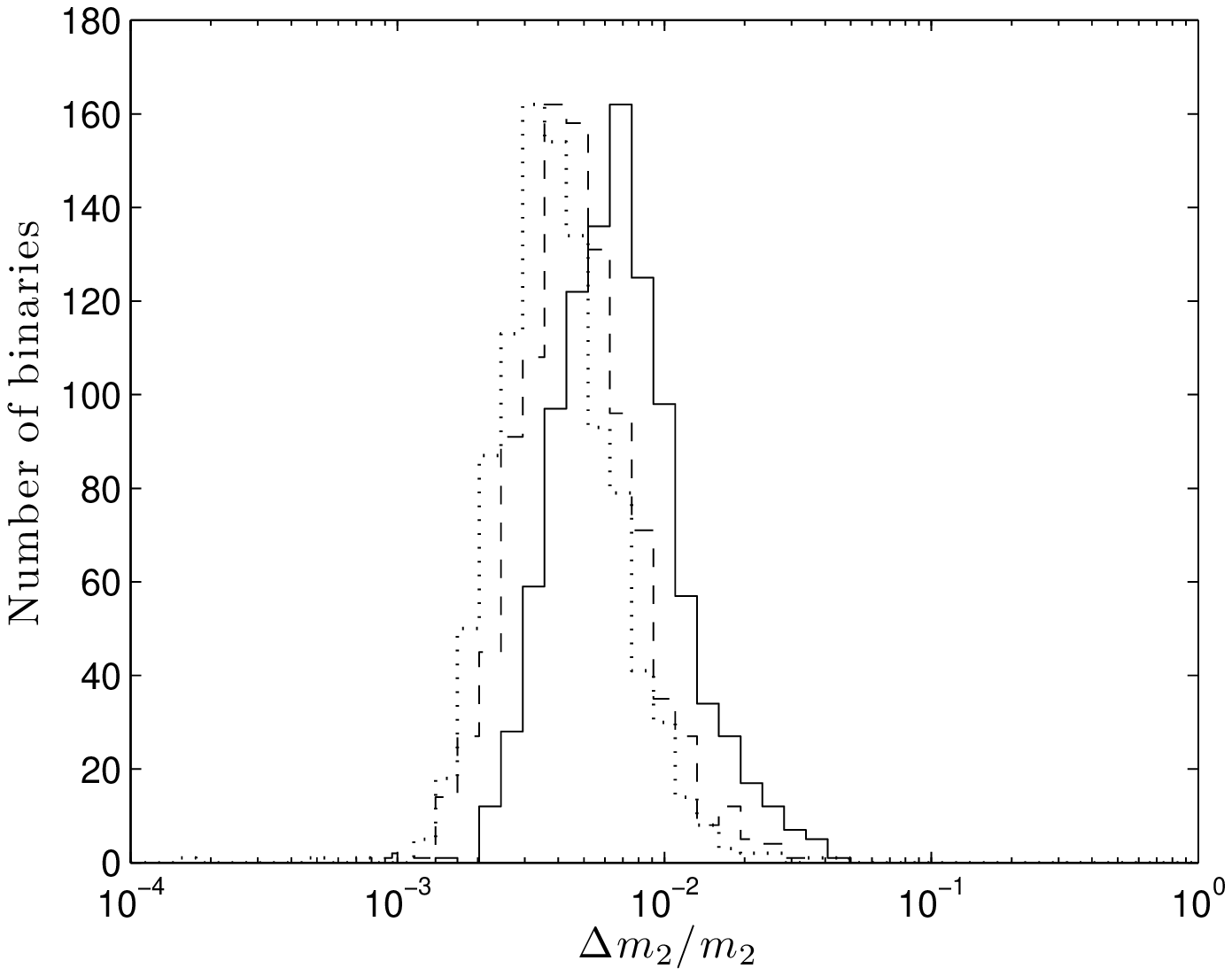}
\caption{Same as Fig. \ref{fig:2anoHH}, but for the fractional error in mass, $\Delta m_2/m_2$.}
\label{fig:m2noHH}
\end{figure}

We now turn to the intrinsic parameters of the system, its masses and
spins.  Figs. \ref{fig:m1noHH} and \ref{fig:m2noHH} show the errors
in the two black hole masses for the three cases we consider.  Medians
of $\Delta m_1/m_1$ are $4.84\times 10^{-3}$ for the random-spin case,
$5.70\times 10^{-3}$ for a system with hot gas, and $8.23\times
10^{-3}$ for a system with cold gas.  For $\Delta m_2/m_2$, these
numbers are $3.84\times 10^{-3}$, $4.54\times 10^{-3}$, and
$6.55\times 10^{-3}$, respectively.  The impact of partially aligned
spins does not seem to be as strong on the masses as on the sky
position; the mass errors change by less than a factor of 2.  In Paper
I, we looked at precession improvements not in individual masses but
in chirp mass and reduced mass, where we saw factors of $\sim 10$ and
$\sim 100-1000$ improvement, respectively.  Clearly, restricting the
spin directions does not remove this entire improvement; even a
limited amount of precession appears to significantly aid mass
determination.  We can check this assertion using our new code by running a case
with only $1^\circ$ alignment between the spins and the orbit.  We find that
the $10^\circ$ results improve on the $1^\circ$ results by a factor of 3.  By contrast, the sky position and distance errors differ by
only $15-20\%$.

Table \ref{table:m1m2noHH} shows the results for different masses.  We
see a much stronger dependence on mass ratio here than for the
extrinsic parameters.  For example, while the cold gas degradation is
less than a factor of 2 for the (roughly) 3:1 mass ratio case considered in
Figs.\ \ref{fig:m1noHH} and \ref{fig:m2noHH}, it reaches a factor of $\sim 9$
for the equal-mass case $m_1 = m_2 = 10^5\ M_\odot$.  This is unusual,
since precession is known, at least for extrinsic parameters, to have
a stronger impact for unequal masses due to increased complexity in
the signal.  It is possible that the lack of this complexity
essentially ``gives away'' that the masses are equal, making them
easier to determine from the extremely well-measured chirp mass.

\begin{figure}[!t]
\includegraphics[scale=0.425]{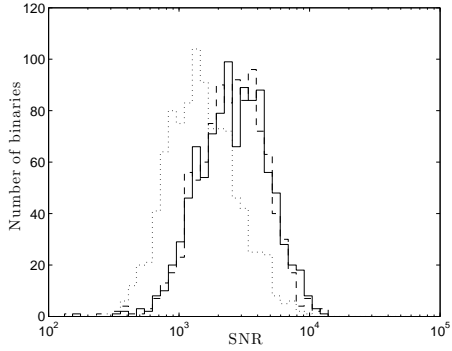}
\caption{Signal-to-noise ratio for binaries with randomly aligned
  spins (dotted line), spins restricted to within $30^\circ$ of the
  orbital angular momentum (dashed line), and spins restricted to
  within $10^\circ$ of the orbital angular momentum (solid line).
  Here $m_1 = 10^6 M_\odot$, $m_2 = 10^5 M_\odot$, and $z = 1$.}
\label{fig:SNRnoHH}
\end{figure}

Interestingly, there are some examples of 10:1 mass ratio systems that
break our general trend; in these cases, we find that partially
aligned spins actually do better than random spins.  This seemingly
counterintuitive result can be explained by our choice of the minimum
energy circular orbit (MECO) as the waveform cutoff.  Binaries with
aligned spins have a smaller MECO (with a corresponding high inspiral
cutoff frequency) and thus accumulate more SNR than those with spins
out of alignment (as many in the random-spin sample will be).  Figure
\ref{fig:SNRnoHH} shows the SNR for all three cases in a 10:1 binary.
We see that the SNR is substantially larger for the partially aligned
cases (medians of 2588 and 2592, for $30^\circ$ and $10^\circ$,
respectively) than the randomly aligned case (median of 1445).  Even
though these binaries precess less, the increase in SNR makes up for
it in parameter estimation.  It is worth noting that this effect could
also be of use in detecting and measuring particularly high-mass binaries.
For randomly chosen spins, such a binary might be mostly or completely out
of the LISA band.  However, if the spins are aligned by interactions with 
gas, the MECO frequency will be pushed into band.

\begin{table*}[!t]
\begin{center}
\begin{tabular}{|c|c||c|c||c|c||c|c|}
\hline
\multirow{2}{*}{$m_1\ (M_\odot)$} & \multirow{2}{*}{$m_2 (M_\odot)$} & \multicolumn{2}{|c||}{No gas} & \multicolumn{2}{|c||}{Hot gas} & \multicolumn{2}{|c|}{Cold gas} \\
& & $\Delta m_1/m_1$ & $\Delta m_2/m_2$ & $\Delta m_1/m_1$ & $\Delta m_2/m_2$ & $\Delta m_1/m_1$ & $\Delta m_2/m_2$ \\
\hline \hline
$10^5$ & $10^5$ & $3.23\times 10^{-3}$ & $3.24\times 10^{-3}$ & $9.84\times 10^{-3}$ & $9.84\times 10^{-3}$ & 0.0284 & 0.0284\\ 
\hline
$3 \times 10^5$ & $10^5$ & $3.02\times 10^{-3}$ & $2.45\times 10^{-3}$ & $3.95\times 10^{-3}$ & $3.20\times 10^{-3}$ & $5.94\times 10^{-3}$ & $4.83\times 10^{-3}$\\
\hline
$3 \times 10^5$ & $3 \times 10^5$ & $4.50\times 10^{-3}$ & $4.50\times 10^{-3}$ & 0.0129 & 0.0128 & 0.0327 & 0.0328\\
\hline
$10^6$ & $10^5$ & $2.72\times 10^{-3}$ & $1.90\times 10^{-3}$ & $1.90\times 10^{-3}$ & $1.32\times 10^{-3}$ & $2.22\times 10^{-3}$ & $1.55\times 10^{-3}$\\
\hline
$10^6$ & $3 \times 10^5$ & $4.84\times 10^{-3}$ & $3.84\times 10^{-3}$ & $5.70\times 10^{-3}$ & $4.54\times 10^{-3}$ & $8.23\times 10^{-3}$ & $6.55\times 10^{-3}$ \\
\hline
$10^6$ & $10^6$ & $8.05\times 10^{-3}$ & $8.05\times 10^{-3}$& 0.0197 & 0.0197 & 0.0475 & 0.0475\\
\hline
$3 \times 10^6$ & $3 \times 10^5$ & $5.81\times 10^{-3}$ & $4.01\times 10^{-3}$ & $4.49\times 10^{-3}$ & $3.07\times 10^{-3}$ & $5.10\times 10^{-3}$ & $3.51\times 10^{-3}$\\
\hline
$3 \times 10^6$ & $10^6$ & 0.0121 & $9.73\times 10^{-3}$ & 0.0165 & 0.0132 & 0.0233 & 0.0187\\
\hline
$3 \times 10^6$ & $3 \times 10^6$ & 0.0239 & 0.0237 & 0.0536 & 0.0533 & 0.109 & 0.109\\
\hline
$10^7$ & $10^6$ & 0.0176 & 0.0118 & 0.0167 & 0.0109 & 0.0205 & 0.0133\\
\hline
$10^7$ & $3 \times 10^6$ & 0.0431 & 0.0336 & 0.0581 & 0.0447 & 0.0924 & 0.0713\\
\hline
$10^7$ & $10^7$ & 0.381 & 0.388 & 0.424 & 0.423 & 0.967 & 0.971\\
\hline 
\end{tabular}
\caption{Same as Table \ref{table:2a2bnoHH}, but for the mass errors $\Delta m_1/m_1$ and $\Delta m_2/m_2$.}
\label{table:m1m2noHH}
\end{center}
\end{table*}

\begin{figure}[!b]
\includegraphics[scale=0.425]{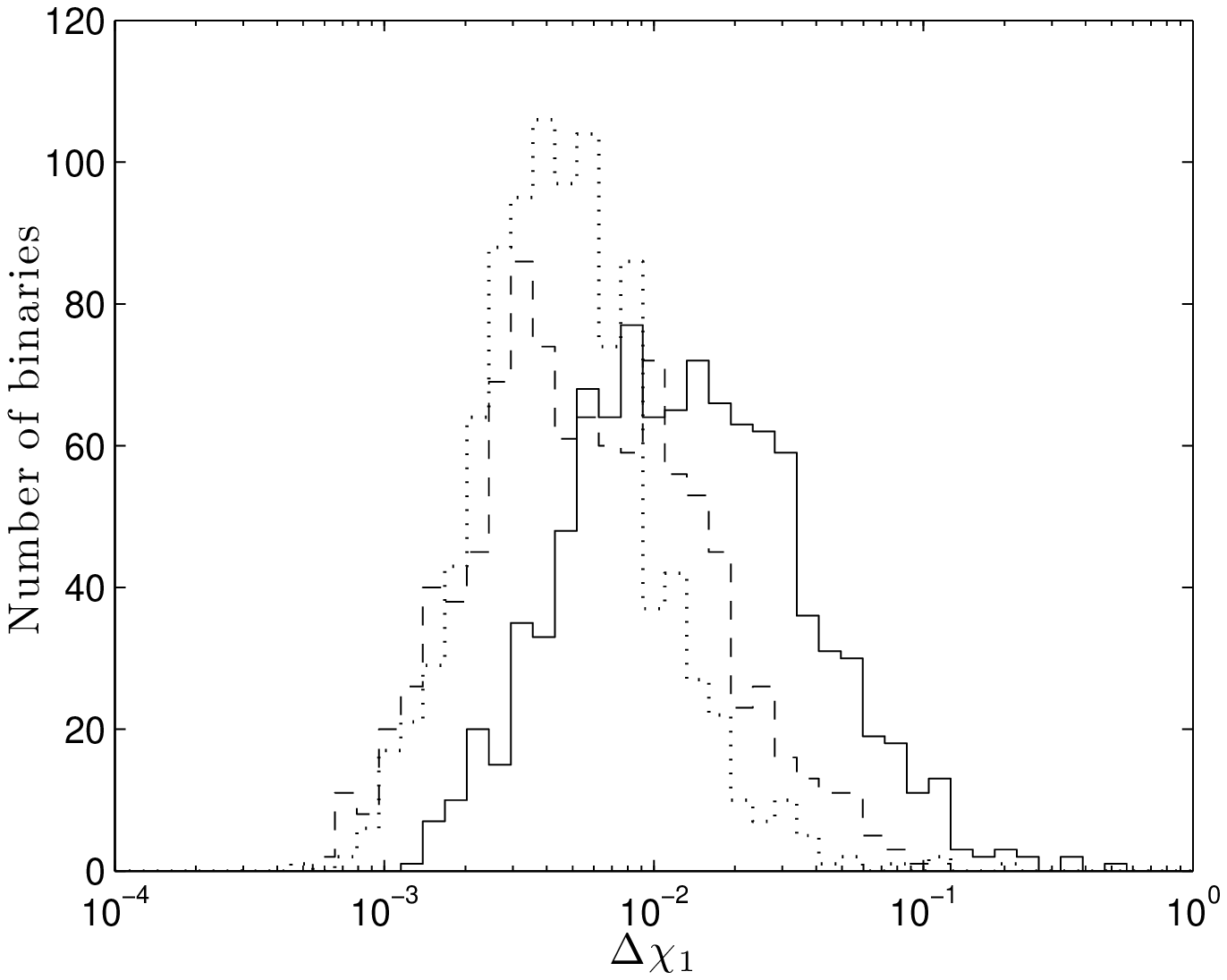}
\caption{Same as Fig. \ref{fig:2anoHH}, but for the error in spin magnitude, $\chi_1 = |\mathbf{S}_1|/m_1^2$.}
\label{fig:chi1noHH}
\end{figure}

\begin{figure}[!t]
\includegraphics[scale=0.425]{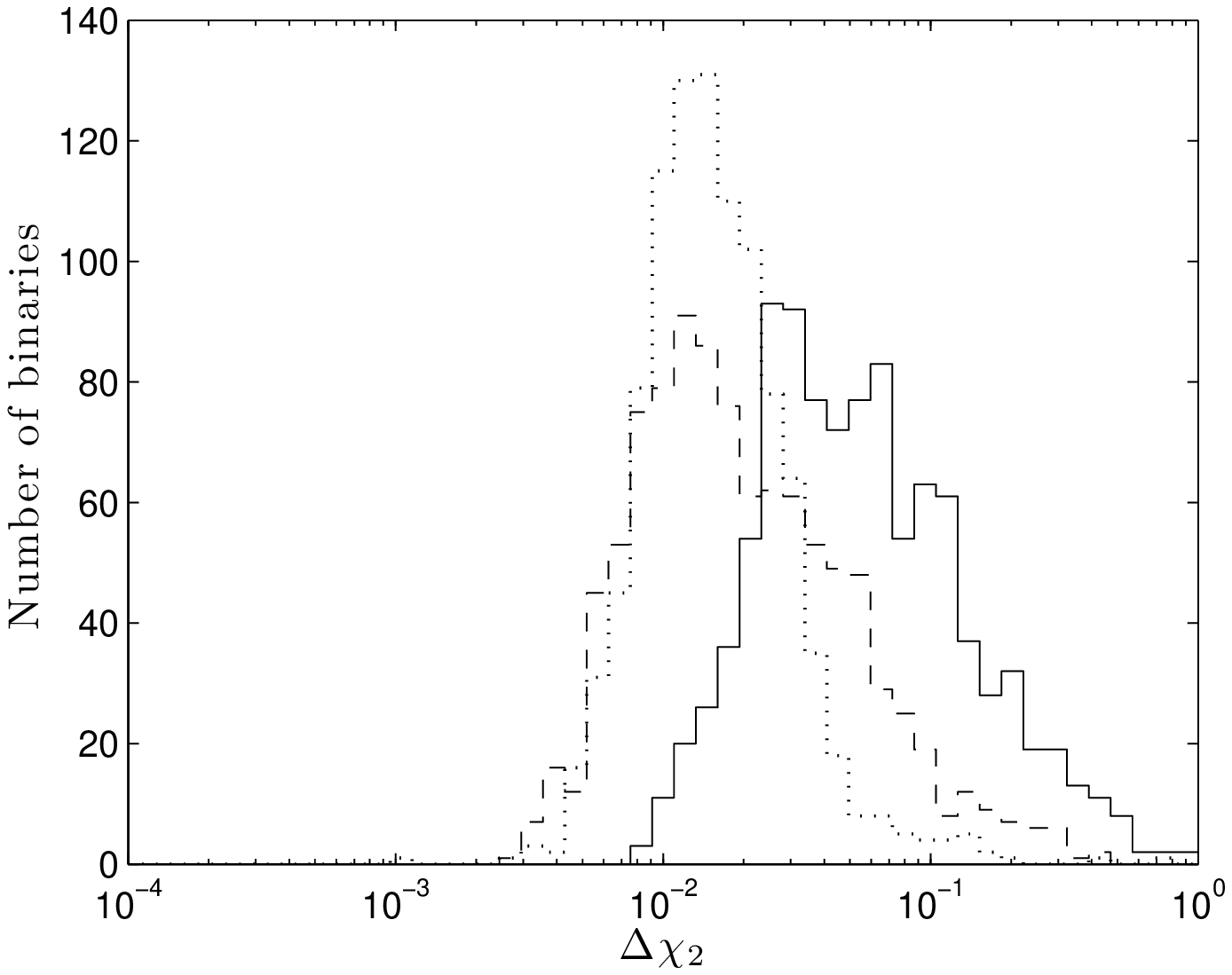}
\caption{Same as Fig. \ref{fig:2anoHH}, but for the error in spin magnitude, $\chi_2 = |\mathbf{S}_2|/m_2^2$.}
\label{fig:chi2noHH}
\end{figure}

Finally, we consider how restriction of spin angles affects
measurement of spin magnitudes; Figs.\ \ref{fig:chi1noHH} and
\ref{fig:chi2noHH} show these results.  For $\chi_1$, the median
varies from $4.55\times 10^{-3}$ for no gas to $5.38\times 10^{-3}$
for hot gas to $1.31\times 10^{-2}$ for cold gas.  For $\chi_2$, the
situation is similar; the medians are $1.48\times 10^{-2}$ (no gas),
$1.72\times 10^{-2}$ (hot gas), and $5.11\times 10^{-2}$ (cold gas).
Although the spin errors are degraded by partial alignment, 
the amount of degradation is somewhat curbed by the contribution of binaries which 
precess away from alignment before MECO.  In addition, the errors at $10^\circ$ alignment are roughly an order of magnitude better than at $1^\circ$ alignment. Similar to the situation with mass measurements, even a small amount of precession can have a huge impact on measuring spin.

\begin{table*}[!t]
\begin{center}
\begin{tabular}{|c|c||c|c||c|c||c|c|}
\hline
\multirow{2}{*}{$m_1\ (M_\odot)$} & \multirow{2}{*}{$m_2 (M_\odot)$} & \multicolumn{2}{|c||}{No gas} & \multicolumn{2}{|c||}{Hot gas} & \multicolumn{2}{|c|}{Cold gas} \\
& & $\Delta \chi_1$ & $\Delta \chi_2$ & $\Delta \chi_1$ & $\Delta \chi_2$ & $\Delta \chi_1$ & $\Delta \chi_2$ \\
\hline \hline
$10^5$ & $10^5$ & 0.0217 & 0.0210 & 0.0311 & 0.0310 & 0.0391 & 0.0388 \\ 
\hline
$3 \times 10^5$ & $10^5$ & $3.18\times 10^{-3}$ & $7.61\times 10^{-3}$ & $4.98\times 10^{-3}$ & 0.0155 & 0.0125 & 0.0467 \\
\hline
$3 \times 10^5$ & $3 \times 10^5$ & 0.0321 & 0.0315 & 0.0390 & 0.0382 & 0.0430 & 0.0417 \\
\hline
$10^6$ & $10^5$ & $1.30\times 10^{-3}$ & 0.0355 & $1.05\times 10^{-3}$ & 0.0225 & $2.22\times 10^{-3}$ & 0.0358\\
\hline
$10^6$ & $3 \times 10^5$ & $4.55\times 10^{-3}$ & 0.0148 & $5.38\times 10^{-3}$ & 0.0172 & 0.0131 & 0.0511 \\
\hline
$10^6$ & $10^6$ & 0.0534 & 0.0505 & 0.0655 & 0.0645 & 0.0774 & 0.0753\\
\hline
$3 \times 10^6$ & $3 \times 10^5$ & $2.38\times 10^{-3}$ & 0.0601 & $1.70\times 10^{-3}$ & 0.0499 & $3.24\times 10^{-3}$ & 0.0659\\
\hline
$3 \times 10^6$ & $10^6$ & $9.92\times 10^{-3}$ & 0.0186 & 0.0111 & 0.0264 & 0.0252 & 0.0787\\
\hline
$3 \times 10^6$ & $3 \times 10^6$ & 0.134 & 0.125 & 0.197 & 0.198 & 0.234 & 0.232\\
\hline
$10^7$ & $10^6$ & $4.86\times 10^{-3}$ & 0.124 & $3.69\times 10^{-3}$ & 0.180 & $7.03\times 10^{-3}$ & 0.231\\
\hline
$10^7$ & $3 \times 10^6$ & 0.0266 & 0.0446 & 0.0298 & 0.0852 & 0.0652 & 0.191\\
\hline
$10^7$ & $10^7$ & 1.69 & 1.54 & 1.28 & 1.31 & 1.86 & 1.87\\
\hline 
\end{tabular}
\caption{Same as Table \ref{table:2a2bnoHH}, but for the spin magnitude errors $\Delta \chi_1$ and $\Delta \chi_2$.}
\label{table:chi1chi2noHH}
\end{center}
\end{table*}

Table \ref{table:chi1chi2noHH} gives spin errors for a broader range
of masses.  Like the mass errors, there is a strong dependence on mass
ratio.  In this case, however, the worst degradation occurs not for
equal masses, but for 3:1 mass ratios, with factors of up to 4
increases in $\Delta \chi_1$ and factors of up to 6 increases in
$\Delta \chi_2$.  The 10:1 cases show some degradation at
$10^\circ$ alignment, but many cases are slightly improved at $30^\circ$ alignment.  As in the case of mass measurements (cf.\ Table
\ref{table:m1m2noHH}), this can be attributed to the increased SNR for
aligned binaries.

\section{Parameter estimation in partially aligned binaries: Including higher harmonics}
\label{sec:results2}

We now move beyond the leading quadrupole waveform to the full
waveforms.  Post-Newtonian corrections to the waveform amplitude are
included up to 2PN order, including both additional quadrupole terms
($\Phi = 2\Phi_\mathrm{orb}$) and subleading (``higher'') harmonics
beyond the quadrupole.  For example, the barycentric waveform $h_+(t)$
can be written up to 1PN order in the amplitude as

\begin{align}
h_+(t) = &\frac{2\mu x}{D_L}\Biggl\{(1+c_i^2)\cos 2\Phi_\mathrm{orb} + x^{1/2}\frac{s_i}{8}\frac{\Delta m}{M}[(5+c_i^2)
\nonumber \\
&\times \cos \Phi_\mathrm{orb}
- 9(1+c_i^2)\cos 3\Phi_\mathrm{orb}] + x \biggl[-\frac{1}{6}(19
\nonumber \\
&+9c_i^2-2c_i^4 - \eta(19-11c_i^2-6c_i^4))\cos 2\Phi_\mathrm{orb}
\nonumber \\
&+ \frac{4}{3}s_i^2(1+c_i^2)(1-3\eta)\cos 4\Phi_\mathrm{orb}\biggr]\Biggr\} \, ,
\label{eq:HH}
\end{align}
where $x = (M\omega)^{2/3}$, $c_i = \cos(\mathbf{\hat{L}}\cdot \mathbf{\hat{n}})$, $s_i =
\sin(\mathbf{\hat{L}}\cdot \mathbf{\hat{n}})$, 
and $\Delta m = m_1-m_2$.  Here we see both extra harmonics and a 1PN
correction to the quadrupole harmonic.  Note that the odd harmonics
only contribute if $m_1 \neq m_2$; just like spin precession, higher
harmonic corrections are more complex for unequal masses.  Further
terms (including the $\times$ polarization) can be found in \cite{b06}
(albeit with some differences in sign convention).

\begin{figure}[!b]
\includegraphics[scale=0.54]{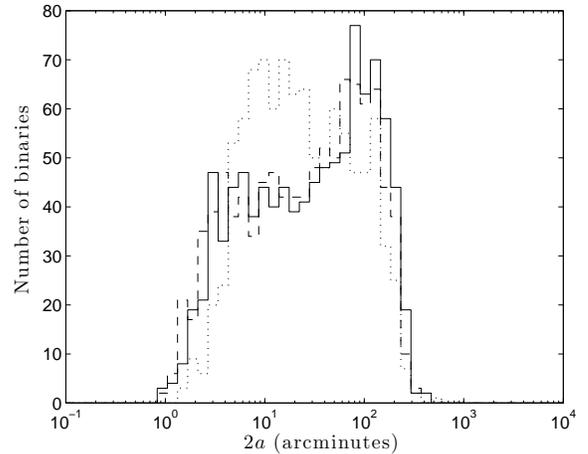}
\caption{Distribution of $2a$, the major axis of the sky position error ellipse, for
  binaries with randomly aligned spins (dotted line), spins restricted
  to within $30^\circ$ of the orbital angular momentum (dashed line),
  and spins restricted to within $10^\circ$ of the orbital angular
  momentum (solid line).  Here $m_1 = 10^6 M_\odot$, $m_2 = 3 \times
  10^5 M_\odot$, and $z = 1$.  Higher harmonics are now included in
  the waveform model.}
\label{fig:2aHH}
\end{figure}

It has been shown that higher harmonic corrections can improve
parameter estimation much like spin precession does \cite{aiss07,
  aissv07, ts08, pc08}.  In the case of higher harmonics, degeneracies
are broken due to the different sky position dependence of each
harmonic.  However, these studies did not include precession and so
could not comment on how the two effects would combine.  More
recently, both effects were included in a parameter estimation study
by Klein {\it et al.}\ (Ref.\ \cite{kjs09}).  Their results demonstrate that
including both precession and higher harmonics improves measurement
accuracy, but, at least for extrinsic variables (sky position and
distance), the combined improvement is not as drastic as the
improvement from each effect on its own.  This indicates that at least
in some cases, precession and higher harmonics encode similar
information.  We might therefore expect that in partially aligned
binaries for which spin precession exists but is suppressed, the
inclusion of higher harmonics may make up for this suppression and
restore much, if not all, of the lost parameter estimation capability.
In this section, we test that expectation.

\begin{figure}[!b]
\includegraphics[scale=0.54]{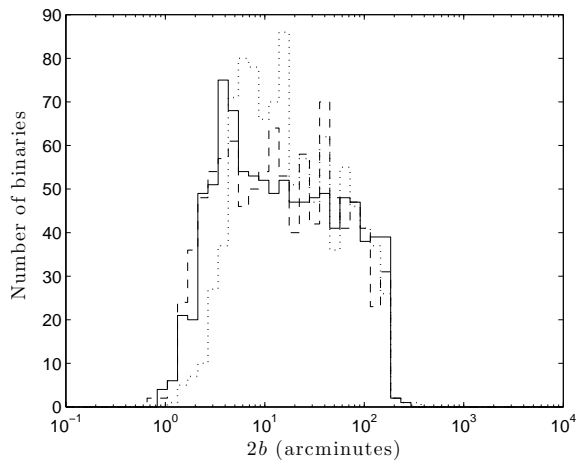}
\caption{Same as Fig. \ref{fig:2aHH}, but for the minor axis $2b$.}
\label{fig:2bHH}
\end{figure}

Figure \ref{fig:2aHH} shows the major axis of the sky position error
ellipse for the same binaries as Fig.\ \ref{fig:2anoHH}, except with
higher harmonics now added to the waveform model.  The median value of
$2a$ for random-spin, gas-free systems is 21.7 arcminutes.  Comparing
to the leading quadrupole waveform value of 34.8 arcminutes, we see
that higher harmonics do indeed add some additional information not
contained in precession.  The difference between the two values is a
factor $\sim 1.6$, consistent with the results of \cite{kjs09}.  For
partially aligned systems, the shape of the plot shows that the higher
harmonics have had an important effect; both partially aligned
histograms look much more like the roughly flat random-spin case than
the strongly peaked histograms shown in Fig.\ \ref{fig:2anoHH}.  The
median is 28.2 arcminutes for hot gas and 32.7 arcminutes for cold
gas.  Both results are great improvements on the leading quadrupole
waveform values (factors $\sim 2.2$ and $2.8$, respectively),
indicating that the inclusion of higher harmonics has indeed ``made
up'' for the loss of some spin precession.  Both results are actually
{\em better} than the leading quadrupole, gas-free result of 34.8
arcminutes.  In this case, a full waveform with a small amount of
precession does better than a leading quadrupole waveform with
potentially significant precession.

Figure \ref{fig:2bHH} shows the results for the minor axis $2b$, with
medians 16.1 arcminutes for random spins and 13.7 arcminutes for
 both $30^\circ$ and $10^\circ$ alignment.
These are all better than the leading quadrupole, random-spin result
of 24.6 arcminutes.  More interestingly, we see that when higher
harmonics are included, the ``cold gas'' and ``hot gas'' errors are
smaller than the ``no gas'' errors.  As discussed in the previous
section, this is due to the improvement in SNR in the aligned case.
Higher harmonics break degeneracies well enough that it is more
beneficial to have partially aligned binaries with more SNR and less
precession than randomly aligned binaries with less SNR and more
precession. 

\begin{table}[!t]
\begin{center}
\begin{tabular}{|c|c||c|c||c|c||c|c|}
\hline
\multirow{2}{*}{$m_1\ (M_\odot)$} & \multirow{2}{*}{$m_2 (M_\odot)$} & \multicolumn{2}{|c||}{No gas} & \multicolumn{2}{|c||}{Hot gas} & \multicolumn{2}{|c|}{Cold gas} \\
& & $2a$ & $2b$ & $2a$ & $2b$ & $2a$ & $2b$ \\
\hline \hline
$10^5$ & $10^5$ & {\bf 21.8} & {\bf 12.3} & 30.3 & {\bf 14.2} & 35.6 & {\bf 15.9} \\ 
\hline
$3 \times 10^5$ & $10^5$ & {\bf 14.3} & {\bf 9.67} & 19.9 & {\bf 10.1} & 31.0 & 13.2\\
\hline
$3 \times 10^5$ & $3 \times 10^5$ & {\bf 26.2} & {\bf 16.2} & {\bf 30.4} & {\bf 18.3} & 39.1 & 25.6\\
\hline
$10^6$ & $10^5$ & {\bf 14.0} & {\bf 11.9} & {\bf 13.7} & {\bf 9.25} & {\bf 17.6} & {\bf 9.69} \\
\hline
$10^6$ & $3 \times 10^5$ & {\bf 21.7} & {\bf 16.1} & {\bf 28.2} & {\bf 13.7} & {\bf 32.7} & {\bf 13.7}\\
\hline
$10^6$ & $10^6$ & {\bf 48.1} & {\bf 32.0} & {\it 60.4} & {\bf 37.1} & {\bf 53.0} & {\bf 32.4}\\
\hline
$3 \times 10^6$ & $3 \times 10^5$ & {\bf 29.1} & {\bf 25.8} & {\bf 25.7} & {\bf 20.0} & {\bf 35.8} & {\bf 24.4}\\
\hline
$3 \times 10^6$ & $10^6$ & {\bf 36.0} & {\bf 26.8} & {\em 48.2} & {\bf 27.4} & 58.2 & {\bf 30.2}\\
\hline
$3 \times 10^6$ & $3 \times 10^6$ & {\bf 63.5} & {\bf 39.8} & 103 & 58.1 & 109 & 66.5\\
\hline
$10^7$ & $10^6$ & {\bf 36.7} & {\bf 32.4} & {\bf 38.9} & {\bf 27.4} & 54.7 & {\bf 31.3}\\
\hline
$10^7$ & $3 \times 10^6$ & {\bf 45.0} & {\bf 32.8} & {\bf 65.1} & {\bf 33.0} & 82.7 & {\bf 36.5}\\
\hline
$10^7$ & $10^7$ & {\bf 114} & {\bf 65.6} & {\bf 144} & {\bf 80.1} & 228 & 115\\
\hline 
\end{tabular}
\caption{Median sky position major axis $2a$ and minor axis $2b$, in
  arcminutes, for binaries of various masses at $z = 1$, in the ``no
  gas'' (random-spin), ``hot gas'' ($30^\circ$ alignment), and ``cold
  gas'' ($10^\circ$ alignment) cases when higher harmonics are
  included in the waveform model.  {\bf Bold} entries are those that
  do better than the no gas case when higher harmonics are ignored
  (i.e. Table \ref{table:2a2bnoHH}).  {\em Italic} entries do worse than
  that case, but only by 10\% or less.}
\label{table:2a2bHH}
\end{center}
\end{table}

Table \ref{table:2a2bHH} shows results for a variety of masses.  All
show improvement from Table \ref{table:2a2bnoHH}; however, the
improvement is not always as strong as in the case discussed above
($m_1 = 10^6\ M_\odot, m_2 = 3\times 10^5\ M_\odot$).  Bold text
indicates cases in which the errors match or improve upon the results
from the leading quadrupole waveform for random spins.  Italics
indicate errors which are worse, but by no more than 10\%.  Because of 
statistical issues, these cases could very well be ``bold'' in a
different Monte Carlo run, so we will consider them as such for
purposes of summarizing the results.  While hot gas ($30^\circ$
alignment) systems achieve this particular benchmark for a majority of
mass cases, cold gas ($10^\circ$ alignment) systems do not.  Cold gas systems
do, however, meet it for a majority of mass cases if only the minor axis is
considered; $2b$ generally fares better than $2a$.  Both axes exhibit cases
where errors are smaller with alignment than without, the minimum sometimes 
occurring in a ``sweet spot'' of $30^\circ$ alignment and sometimes at $10^\circ$ alignment.
In general, errors are better for larger mass
ratios.  This is to be expected because both
higher harmonics and precession have a more complicated structure for
larger mass ratios.  Finally, the improvements are worst for the
smallest masses, where the higher harmonics (except the
$\cos\Phi_\mathrm{orb}$ terms, which are technically ``lower''
harmonics) begin to go out of band.
\begin{figure}[!t]
\includegraphics[scale=0.54]{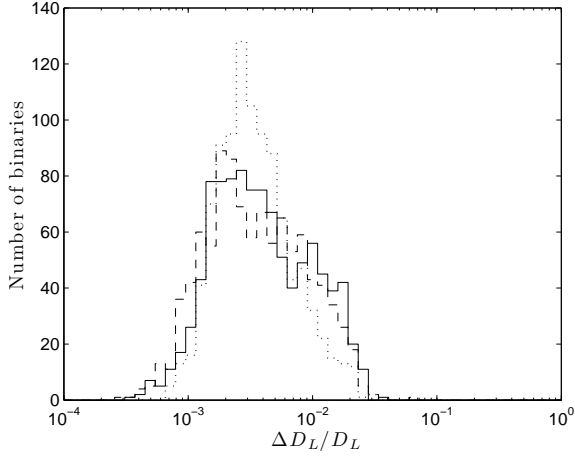}
\caption{Same as Fig. \ref{fig:2aHH}, but for the fractional error in luminosity distance, $\Delta D_L/D_L$.}
\label{fig:DHH}

\end{figure}

\begin{table}[!b]
\begin{center}
\begin{tabular}{|c|c||c|c|c|}
\hline
$m_1\ (M_\odot)$ & $m_2 (M_\odot)$ & No gas & Hot gas & Cold gas \\
\hline \hline
$10^5$ & $10^5$ & $\mathbf{3.83\times 10^{-3}}$ & $5.95\times 10^{-3}$ & $7.23\times 10^{-3}$\\ 
\hline
$3 \times 10^5$ & $10^5$ & $\mathbf{1.89\times 10^{-3}}$ & $2.88\times 10^{-3}$ & $4.05\times 10^{-3}$ \\
\hline
$3 \times 10^5$ & $3 \times 10^5$ & $\mathbf{4.16\times 10^{-3}}$ & $\mathbf{4.78\times 10^{-3}}$ & $\mathbf{5.45\times 10^{-3}}$\\
\hline
$10^6$ & $10^5$ & $\mathbf{2.07\times 10^{-3}}$ & $\mathbf{1.66\times 10^{-3}}$ & $\mathbf{2.12\times 10^{-3}}$\\
\hline
$10^6$ & $3 \times 10^5$ & $\mathbf{3.20\times 10^{-3}}$ & $\mathbf{3.20\times 10^{-3}}$ & $\mathbf{3.54\times 10^{-3}}$\\
\hline
$10^6$ & $10^6$ & $\mathbf{7.16\times 10^{-3}}$ & $\mathbf{7.62\times 10^{-3}}$ & $\mathbf{7.48\times 10^{-3}}$\\
\hline
$3 \times 10^6$ & $3 \times 10^5$ & $\mathbf{4.01\times 10^{-3}}$ & $\mathbf{3.29\times 10^{-3}}$ & $\mathbf{3.77\times 10^{-3}}$ \\
\hline
$3 \times 10^6$ & $10^6$ & $\mathbf{5.38\times 10^{-3}}$ & $\mathbf{6.23\times 10^{-3}}$ & $\mathbf{6.79\times 10^{-3}}$ \\
\hline
$3 \times 10^6$ & $3 \times 10^6$ & {\bf 0.0115} & {\it 0.0139} & 0.0152\\
\hline
$10^7$ & $10^6$ & $\mathbf{5.69\times 10^{-3}}$ & $\mathbf{5.69\times 10^{-3}}$ & $\mathbf{7.42\times 10^{-3}}$ \\
\hline
$10^7$ & $3 \times 10^6$ & $\mathbf{7.69\times 10^{-3}}$ & $\mathbf{9.02\times 10^{-3}}$ & {\bf 0.0121}\\
\hline
$10^7$ & $10^7$ & {\bf 0.0223} & {\bf 0.0270} & {\bf 0.0302}\\
\hline 
\end{tabular}
\caption{Same as Table \ref{table:2a2bHH}, but for the fractional error in luminosity distance, $\Delta D_L/D_L$.}
\label{table:DHH}
\end{center}
\end{table}

Figure \ref{fig:DHH} shows the results for luminosity distance errors
$\Delta D_L/D_L$.  Here the medians are $3.20\times 10^{-3}$,
$3.20\times 10^{-3}$, and $3.54\times 10^{-3}$ for the no gas, hot
gas, and cold gas cases, respectively.  Again, these are all better
than the leading quadrupole, no gas value of $5.24\times 10^{-3}$, as
we might expect since distance determination is strongly tied to sky
position determination.  Table \ref{table:DHH} gives the results for
various masses.  Most cases beat the leading quadrupole, no gas values
of Table \ref{table:DnoHH} or come within 10\%.  That is, except for
the lowest mass systems, using the full waveform essentially always
brings the distance errors for aligned spins back to the level of
random spins.  In this respect, distance errors are similar to (and
even a bit better than) the minor axis of the sky position error
ellipse.  Finally, as with sky position, some mass cases feature errors which 
decrease as spins become aligned.

\begin{figure}[!t]
\includegraphics[scale=0.54]{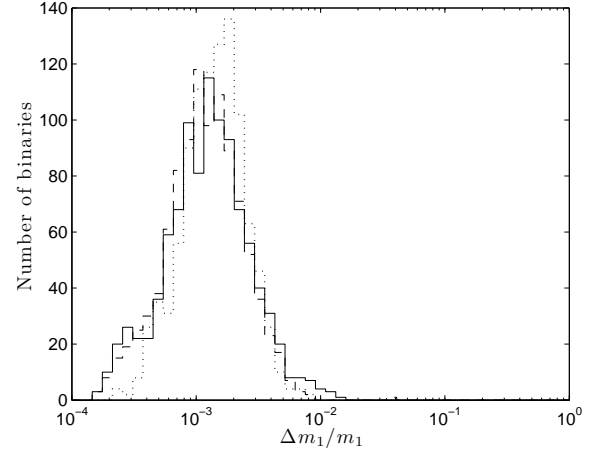}
\caption{Same as Fig. \ref{fig:2aHH}, but for the fractional error in mass, $\Delta m_1/m_1$.}
\label{fig:m1HH}
\end{figure}

\begin{figure}[!b]
\includegraphics[scale=0.54]{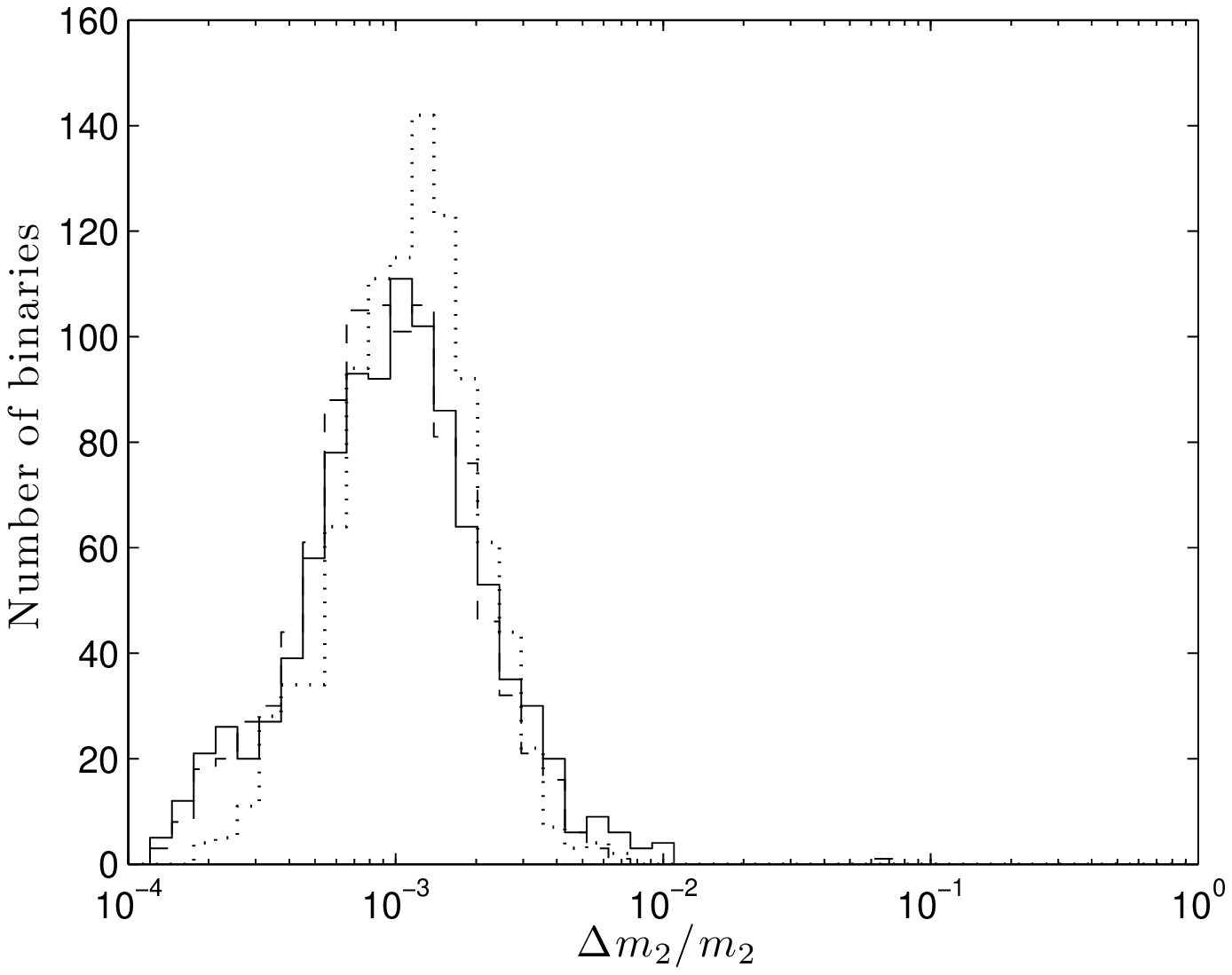}
\caption{Same as Fig. \ref{fig:2aHH}, but for the fractional error in mass, $\Delta m_2/m_2$.}
\label{fig:m2HH}
\end{figure}

\begin{table*}[!th]
\begin{center}
\begin{tabular}{|c|c||c|c||c|c||c|c|}
\hline
\multirow{2}{*}{$m_1\ (M_\odot)$} & \multirow{2}{*}{$m_2 (M_\odot)$} & \multicolumn{2}{|c||}{No gas} & \multicolumn{2}{|c||}{Hot gas} & \multicolumn{2}{|c|}{Cold gas} \\
& & $\Delta m_1/m_1$ & $\Delta m_2/m_2$ & $\Delta m_1/m_1$ & $\Delta m_2/m_2$ & $\Delta m_1/m_1$ & $\Delta m_2/m_2$ \\
\hline \hline
$10^5$ & $10^5$ & $\mathbf{1.22\times 10^{-3}}$ & $\mathbf{1.21\times 10^{-3}}$ & $\mathbf{1.16\times 10^{-3}}$ & $\mathbf{1.16\times 10^{-3}}$ & $\mathbf{1.24\times 10^{-3}}$ & $\mathbf{1.24\times 10^{-3}}$ \\ 
\hline
$3 \times 10^5$ & $10^5$ & $\mathbf{1.29\times 10^{-3}}$ & $\mathbf{1.04\times 10^{-3}}$ & $\mathbf{1.43\times 10^{-3}}$ & $\mathbf{1.16\times 10^{-3}}$ & $\mathbf{1.77\times 10^{-3}}$ & $\mathbf{1.44\times 10^{-3}}$\\
\hline
$3 \times 10^5$ & $3 \times 10^5$ & $\mathbf{6.14\times 10^{-4}}$ & $\mathbf{6.13\times 10^{-4}}$ & $\mathbf{5.37\times 10^{-4}}$ & $\mathbf{5.35\times 10^{-4}}$ & $\mathbf{5.12\times 10^{-4}}$ & $\mathbf{5.14\times 10^{-4}}$\\
\hline
$10^6$ & $10^5$ & $\mathbf{1.32\times 10^{-3}}$ & $\mathbf{9.19\times 10^{-4}}$ & $\mathbf{9.23\times 10^{-4}}$ & $\mathbf{6.39\times 10^{-4}}$ & $\mathbf{1.10\times 10^{-3}}$ & $\mathbf{7.64\times 10^{-4}}$  \\
\hline
$10^6$ & $3 \times 10^5$ & $\mathbf{1.44\times 10^{-3}}$ & $\mathbf{1.15\times 10^{-3}}$ & $\mathbf{1.17\times 10^{-3}}$ & $\mathbf{9.32\times 10^{-4}}$ & $\mathbf{1.24\times 10^{-3}}$ & $\mathbf{9.85\times 10^{-4}}$\\
\hline
$10^6$ & $10^6$ & $\mathbf{1.27\times 10^{-3}}$ & $\mathbf{1.28\times 10^{-3}}$ & $\mathbf{7.64\times 10^{-4}}$ & $\mathbf{7.65\times 10^{-4}}$ & $\mathbf{7.25\times 10^{-4}}$ & $\mathbf{7.21\times 10^{-4}}$\\
\hline
$3 \times 10^6$ & $3 \times 10^5$ & $\mathbf{2.50\times 10^{-3}}$ & $\mathbf{1.73\times 10^{-3}}$ & $\mathbf{1.47\times 10^{-3}}$ & $\mathbf{1.00\times 10^{-3}}$ & $\mathbf{1.62\times 10^{-3}}$ & $\mathbf{1.12\times 10^{-3}}$\\
\hline
$3 \times 10^6$ & $10^6$ & $\mathbf{3.06\times 10^{-3}}$ & $\mathbf{2.45\times 10^{-3}}$ & $\mathbf{2.55\times 10^{-3}}$ & $\mathbf{2.05\times 10^{-3}}$ & $\mathbf{2.65\times 10^{-3}}$ & $\mathbf{2.13\times 10^{-3}}$\\
\hline
$3 \times 10^6$ & $3 \times 10^6$ & $\mathbf{2.04\times 10^{-3}}$ & $\mathbf{2.03\times 10^{-3}}$ & $\mathbf{1.48\times 10^{-3}}$ & $\mathbf{1.48\times 10^{-3}}$ & $\mathbf{1.47\times 10^{-3}}$ & $\mathbf{1.47\times 10^{-3}}$\\
\hline
$10^7$ & $10^6$ & $\mathbf{4.52\times 10^{-3}}$ & $\mathbf{3.04\times 10^{-3}}$ & $\mathbf{3.17\times 10^{-3}}$ & $\mathbf{2.08\times 10^{-3}}$ & $\mathbf{3.55\times 10^{-3}}$ & $\mathbf{2.35\times 10^{-3}}$\\
\hline
$10^7$ & $3 \times 10^6$ & $\mathbf{4.04\times 10^{-3}}$ & $\mathbf{3.12\times 10^{-3}}$ & $\mathbf{3.56\times 10^{-3}}$ & $\mathbf{2.73\times 10^{-3}}$ & $\mathbf{4.21\times 10^{-3}}$ & $\mathbf{3.28\times 10^{-3}}$\\
\hline
$10^7$ & $10^7$ & $\mathbf{5.05\times 10^{-3}}$ & $\mathbf{4.90\times 10^{-3}}$ & $\mathbf{4.17\times 10^{-3}}$ & $\mathbf{4.01\times 10^{-3}}$ & $\mathbf{4.50\times 10^{-3}}$ & $\mathbf{4.18\times 10^{-3}}$\\
\hline 
\end{tabular}
\caption{Same as Table \ref{table:2a2bHH}, but for the mass errors $\Delta m_1/m_1$ and $\Delta m_2/m_2$.}
\label{table:m1m2HH}
\end{center}
\end{table*}

Figs. \ref{fig:m1HH} and \ref{fig:m2HH} show the results for masses
$m_1$ and $m_2$.  The medians for $\Delta m_1/m_1$ are $1.44\times
10^{-3}$ (no gas), $1.17\times 10^{-3}$ (hot gas), and $1.24\times
10^{-3}$ (cold gas).  For $\Delta m_2/m_2$, these are $1.15\times
10^{-3}$ (no gas), $9.32\times 10^{-4}$ (hot gas), and $9.85\times
10^{-4}$ (cold gas).  These are all significant improvements on the
leading quadrupole, no gas case, about a factor of 3-4.  It seems that
for mass errors, higher harmonics are more useful than spin
precession.

\begin{figure}[!b]
\includegraphics[scale=0.54]{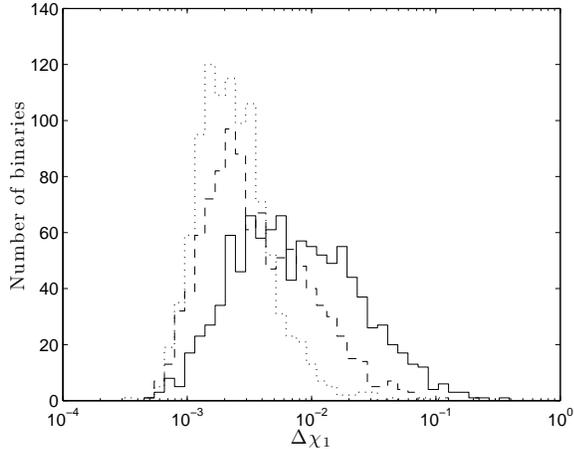}
\caption{Same as Fig. \ref{fig:2aHH}, but for the error in spin
  magnitude, $\chi_1 = |\mathbf{S}_1|/m_1^2$.}
\label{fig:chi1HH}
\end{figure}

This conclusion is supported by Table \ref{table:m1m2HH}.  Here we see
that every case is significantly improved over the leading quadrupole,
random-spin result.  The effect is strongest at higher masses, where
the improvement can be an order of magnitude or more.  This is due to
the well-known effect of higher harmonics on higher mass signals:
Normally these signals are only in band for a short amount of time.
The inclusion of higher frequencies keeps the signal in band longer,
allowing for the accumulation of more phase and better mass
determination.  The improvement is also greater for equal masses,
similar to what was seen in Sec. \ref{sec:results1} for precession.

While the extrinsic parameter errors were only occasionally reduced by partial 
alignment, this phenomenon occurs almost always for mass errors.  This might be expected, since the effect showed up previously for mass errors (in Table
\ref{table:m1m2noHH}) even without higher harmonics to help break
degeneracies.  In general, the
difference in mass accuracy between gas environments is relatively small.
Partial alignment of spins does not affect mass determination as long
as the signal model includes higher harmonics; such alignment may even
help measure mass, at least slightly.

\begin{figure}[!b]
\includegraphics[scale=0.54]{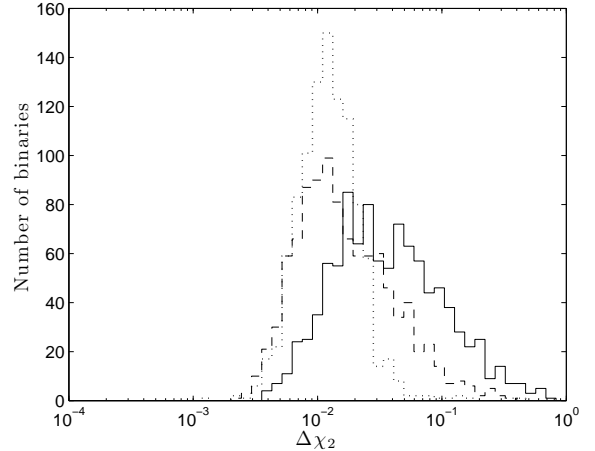}
\caption{Same as Fig. \ref{fig:2aHH}, but for the error in spin
  magnitude, $\chi_2 = |\mathbf{S}_2|/m_2^2$.}
\label{fig:chi2HH}
\end{figure}

\begin{table*}[!p]
\begin{center}
\begin{tabular}{|c|c||c|c||c|c||c|c|}
\hline
\multirow{2}{*}{$m_1\ (M_\odot)$} & \multirow{2}{*}{$m_2 (M_\odot)$} & \multicolumn{2}{|c||}{No gas} & \multicolumn{2}{|c||}{Hot gas} & \multicolumn{2}{|c|}{Cold gas} \\
& & $\Delta \chi_1$ & $\Delta \chi_2$ & $\Delta \chi_1$ & $\Delta \chi_2$ & $\Delta \chi_1$ & $\Delta \chi_2$ \\
\hline \hline
$10^5$ & $10^5$ & {\bf 0.0155} & {\bf 0.0157} & {\bf 0.0178} & {\bf 0.0178} & 0.0286 & 0.0280 \\ 
\hline
$3 \times 10^5$ & $10^5$ & $\mathbf{1.79\times 10^{-3}}$ & $\mathbf{6.18\times 10^{-3}}$ & $\mathbf{3.09\times 10^{-3}}$ & 0.0118 & $7.03\times 10^{-3}$ & 0.0275 \\
\hline
$3 \times 10^5$ & $3 \times 10^5$ & {\bf 0.0207} & {\bf 0.0211} & {\bf 0.0247} & {\bf 0.0243} & {\it 0.0335} & {\it 0.0335}\\
\hline
$10^6$ & $10^5$ & $\mathbf{6.43\times 10^{-4}}$ & {\bf 0.0196} & $\mathbf{6.21\times 10^{-4}}$ & {\bf 0.0137} & $1.61\times 10^{-3}$ & {\bf 0.0263}\\
\hline
$10^6$ & $3 \times 10^5$ & $\mathbf{2.20\times 10^{-3}}$ & {\bf 0.0121} & $\mathbf{3.01\times 10^{-3}}$ & {\bf 0.0145} & $6.88\times 10^{-3}$ & 0.0333\\
\hline
$10^6$ & $10^6$ & {\bf 0.0311} & {\bf 0.0326} & {\bf 0.0385} & {\bf 0.0396} & 0.0603 & 0.0590\\
\hline
$3 \times 10^6$ & $3 \times 10^5$ & $\mathbf{1.07\times 10^{-3}}$ & {\bf 0.0373}& $\mathbf{8.64\times 10^{-4}}$ & {\bf 0.0218} & $\mathbf{2.24\times 10^{-3}}$ & {\bf 0.0408}\\
\hline
$3 \times 10^6$ & $10^6$ & $\mathbf{3.90\times 10^{-3}}$ & {\bf 0.0153} & $\mathbf{5.29\times 10^{-3}}$ & 0.0221 & 0.0129 & 0.0551\\
\hline
$3 \times 10^6$ & $3 \times 10^6$ & {\bf 0.0664} & {\bf 0.0669} & {\bf 0.0932} & {\bf 0.0899} & 0.172 & 0.171\\
\hline
$10^7$ & $10^6$ & $\mathbf{1.75\times 10^{-3}}$ & {\bf 0.0552} & $\mathbf{1.67\times 10^{-3}}$ & {\bf 0.0482} & $\mathbf{4.03\times 10^{-3}}$ & {\bf 0.0872}\\
\hline
$10^7$ & $3 \times 10^6$ & $\mathbf{6.03\times 10^{-3}}$ & {\bf 0.0295} & {\bf 0.0111} & 0.0557 & {\bf 0.0264} & 0.140\\
\hline
$10^7$ & $10^7$ & {\bf 0.495} & {\bf 0.525} & {\bf 0.548} & {\bf 0.581} & {\bf 1.08} & {\bf 1.08}\\
\hline 
\end{tabular}
\caption{Same as Table \ref{table:2a2bHH}, but for the spin magnitude errors $\Delta \chi_1$ and $\Delta \chi_2$.}
\label{table:chi1chi2HH}
\end{center}
\end{table*}

\begin{figure*}[!p]
\includegraphics[scale=0.54]{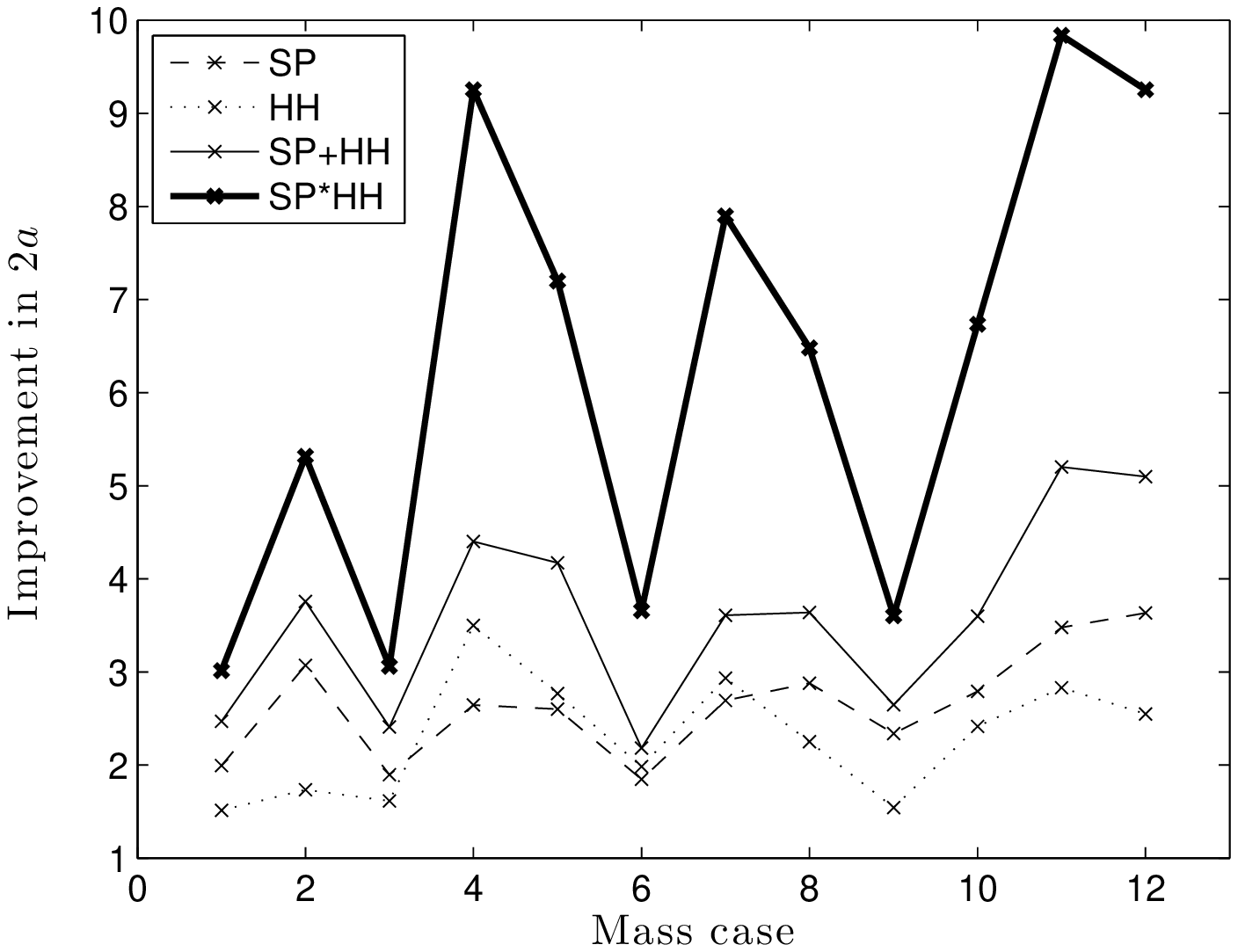}
\includegraphics[scale=0.54]{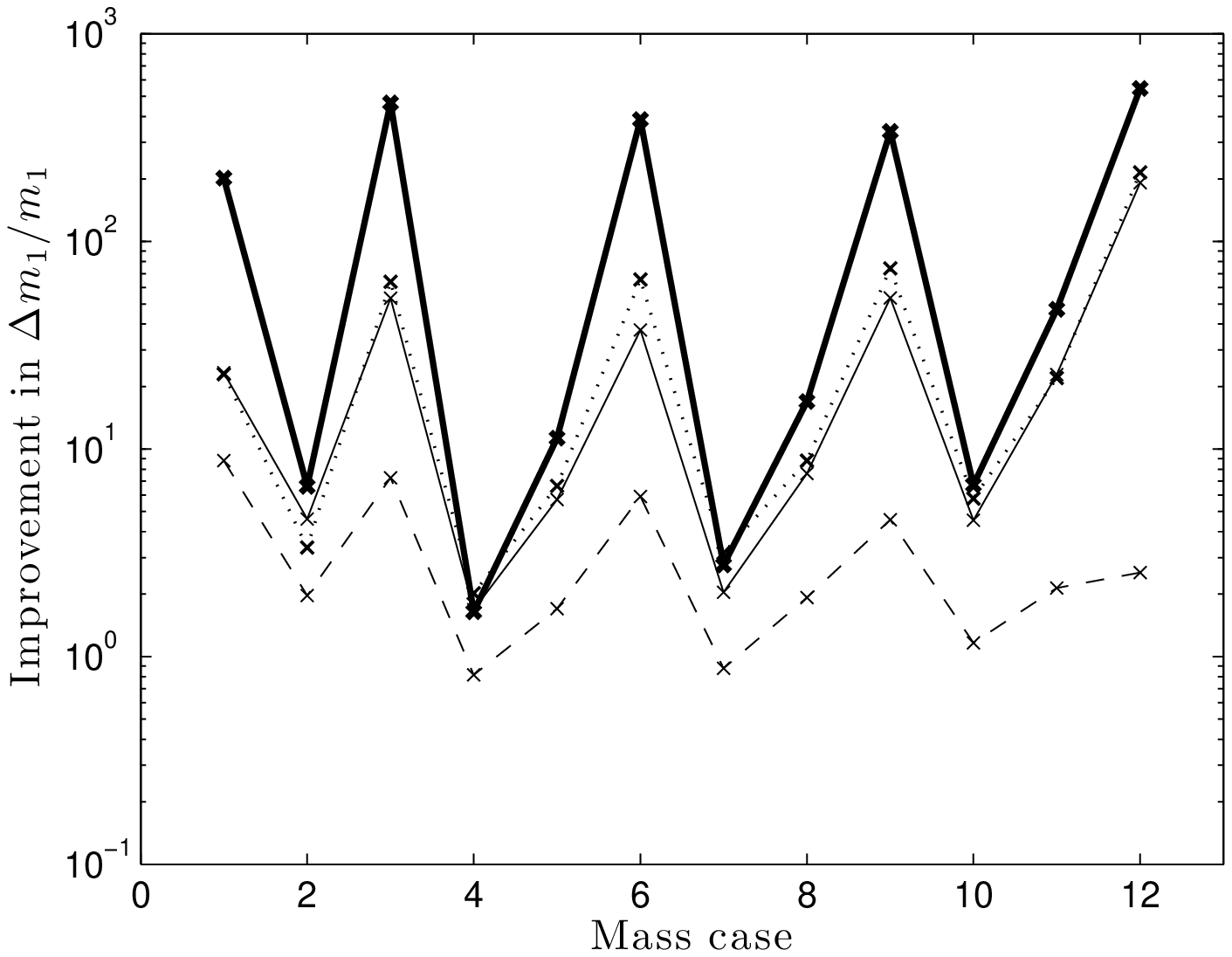}
\includegraphics[scale=0.54]{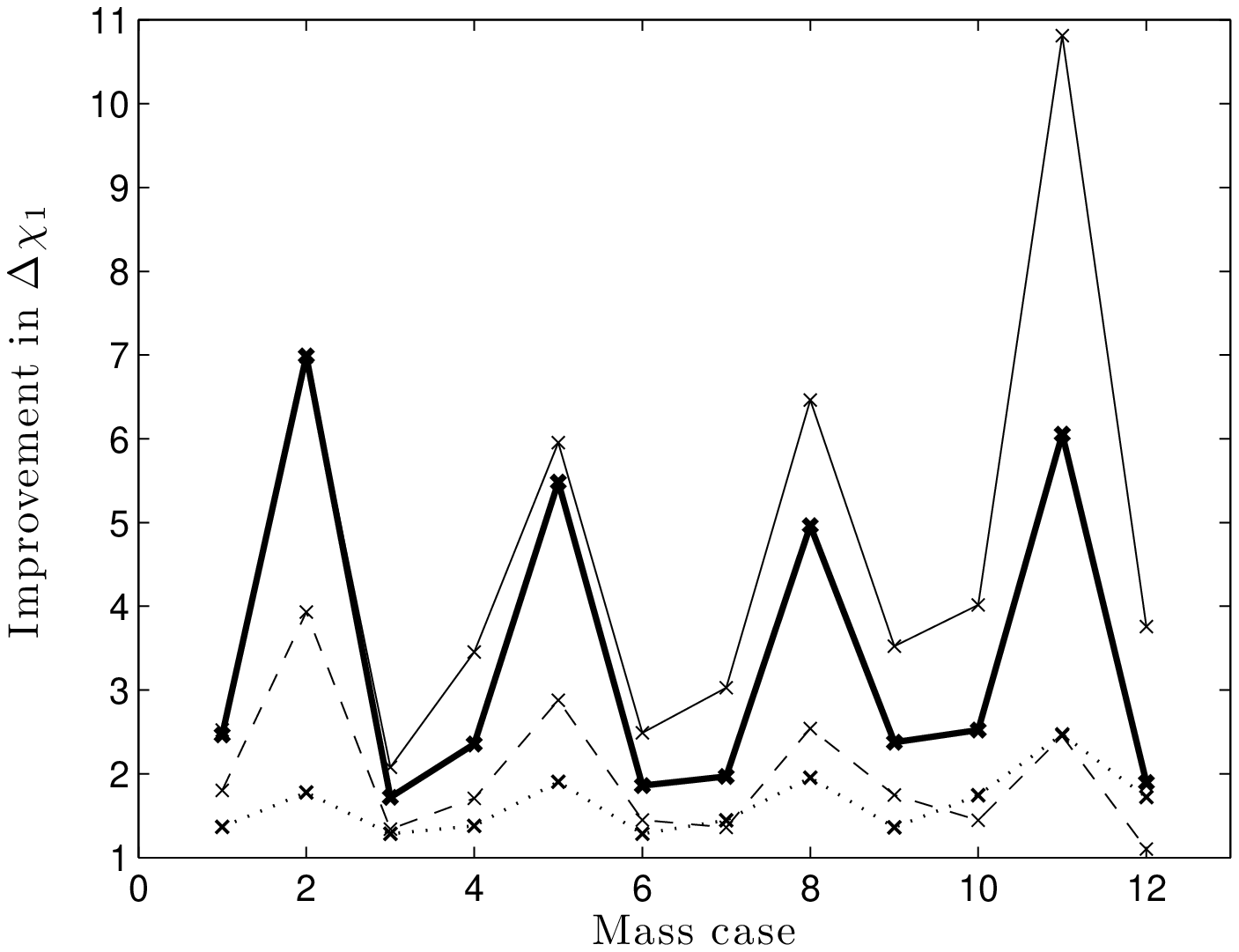}
\includegraphics[scale=0.54]{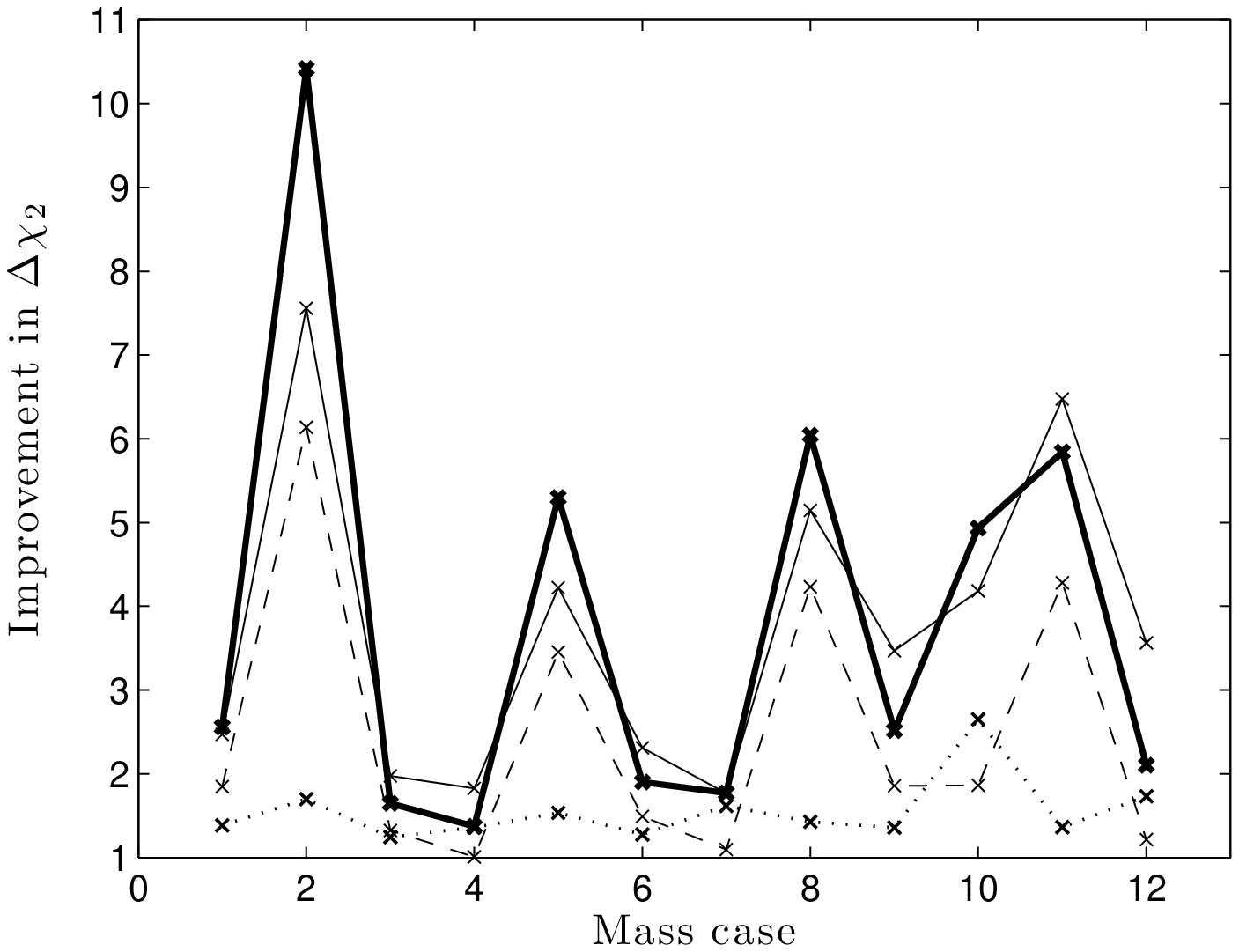}
\caption{Factors by which measurement accuracy improves for different
  parameters when various degeneracy-breaking effects are included in
  the signal: spin precession (SP), higher harmonics (HH), and both
  (SP+HH).  We also show the product of the individual precession and
  harmonic improvements (SP*HH); this represents the naive limit by
  which the two effects would improve measurement accuracy if their
  individual improvements simply combined.  Each point represents one
  of the 12 mass cases, arranged in order from left to right as they
  read top to bottom in Tables
  \ref{table:2a2bnoHH}, \ref{table:DnoHH}, \ref{table:m1m2noHH}, \ref{table:chi1chi2noHH}, \ref{table:2a2bHH}, \ref{table:DHH}, \ref{table:m1m2HH}, and \ref{table:chi1chi2HH}.}
\label{fig:improvements}
\end{figure*}

Finally, Figs.\ \ref{fig:chi1HH} and \ref{fig:chi2HH} present the
errors in spin magnitude.  These figures clearly show that higher
harmonics do {\em not} help spin errors as much as spin precession
does.  This is to be expected, as the spin magnitudes drive the
precession but do not appear in the higher harmonic amplitudes.  Any
gain in spin accuracy due to higher harmonics is a result of
improvement in other parameters (such as the masses) which are
correlated with the spins.  For $\chi_1$, the median errors are
$2.20\times 10^{-3}$ for no gas, $3.01\times 10^{-3}$ for hot gas, and
$6.88\times 10^{-3}$ for cold gas; for $\chi_2$, these numbers are
$1.21\times 10^{-2}$, $1.45\times 10^{-2}$, and $3.33\times 10^{-2}$.
While these errors represent improvements (up to a factor of $\sim 2$) 
over their leading quadrupole waveform counterparts,
it is worth noting that for this ``fiducial'' case, spins are the only
parameters for which the cold gas errors with higher harmonics do not
improve upon the no gas errors for the leading quadrupole waveform.
(Recall, however, that for other masses, sky position and distance
errors do not always achieve this benchmark.)   Table \ref{table:chi1chi2HH} shows the results for various masses.  The
cold gas error only beats the no gas, leading quadrupole error in a few cases.

We conclude this section by looking more directly at the impact of
different degeneracy-breaking effects on parameter errors.  Figure
\ref{fig:improvements} shows multiplicative improvement factors (i.e.,
ratios of errors) when either precession, harmonics, or both effects
are included in the waveform.  For this purpose, we consider binaries
with $10^\circ$ alignment to represent ``no precession'' and those with
random spins to represent ``precession.''  (Remember, though, that even $10^\circ$ alignment permits enough precession to significantly impact mass and spin estimation.)  We also show the product of
the individual improvements from precession and harmonics.  This naive
limit describes how the improvements would combine if each effect were
completely uncorrelated from the other.

For extrinsic parameters (represented in the figure by $2a$; results
for $2b$ and $D_L$ are very similar), the results confirm what was
known previously.  While including both precession and harmonics
improves errors more than one effect alone, the total improvement
falls well short of the naive expectation.  In essence, a degeneracy
can only be broken once.  For mass (shown on a log scale), the same is true, but with the
special feature pointed out above: Once harmonics are included, they
essentially dominate mass accuracy determination.  Spin precession is
then a small liability, because the associated misaligned spins reduce
SNR.  For spin errors, on the other hand, the combined improvement due
to both effects {\it does} roughly match the naive expectation.  In
the case of $\chi_1$, the improvement is actually greater than the
product of the individual improvements.  Different behavior for spin
magnitudes is to be expected, since the information about spin
contained in harmonics is very indirect.  This information is
therefore independent of any information derived directly from
precession.

\section{Conclusions}
\label{sec:conclusions}

Past work has shown that both spin precession and higher harmonics
improve LISA's ability to measure the parameters of merging massive
black hole binaries.  Though these two effects produce similar
degeneracy-breaking effects and similar improvements to measurement
errors, there is one key difference between the two: Higher harmonics
are {\em always} present in the signal (although the strength of odd
harmonics depends on mass ratio).  Spin precession, on the other hand,
may be highly attenuated for {\em physical} reasons, namely, the
partial alignment of spins due to interaction with gas.  In this
paper, we have studied how this partial alignment affects parameter
measurement errors.

Initially ignoring the impact of higher harmonics, we found that sky
position and distance are measured a factor of $\sim 1.5-2$ less
accurately for systems aligned within $30^\circ$ (due to hot gas) and
$\sim 2-3$ less accurately for systems aligned within $10^\circ$ (due
to cold gas).  A degradation of $\sim 3$ would
correspond to an order of magnitude decrease in LISA's ability to
localize a source on the sky and a half order of magnitude decrease in
the ability to localize it in redshift space.  Since systems with gas,
whether hot or cold, are the most likely to produce electromagnetic
counterparts, this means that the results of Paper I and \cite{lh08,lh09}
strongly overestimate our ability to find these counterparts.  Mass
and spin measurements are also degraded by spin alignment, in some
cases by factors up to $\sim 9$.  However, because the masses and spins are already measured quite well, the degradation is not as harmful.

Adding higher harmonics to the signal substantially improves these
results.  In some cases, measurement errors for aligned systems can be
brought below the error level for random spin orientations without
higher harmonics.  For the mass measurements, this improvement happens
in every case.  The minor axis of the sky position error ellipse and
luminosity distance achieve this benchmark less often, though still in
the majority of cases, while the major axis and the spin magnitudes do
not fare as well.  Sometimes parameters 
are actually determined a bit better for aligned spins than
for random ones thanks to the increased SNR measured for aligned spin
systems.  For mass measurements, this happens in almost every case we 
considered.

Although studies like \cite{kjs09} and this one are starting to
finalize expectations for LISA's parameter estimation capabilities,
several avenues of research into the problem still remain.  First is
the issue of including proper astrophysical information, such as the
possibility of partially aligned spins, when analyzing LISA science.
Another example is the recent realization that eccentricity may need
to be included in the waveform model (with a first analysis by Key and
Cornish \cite{kc11}).  In order to make reliable estimates of LISA's
science capabilities, future studies must continue to incorporate the
newest astrophysical developments.  It will also be useful to turn the
problem around and ask what LISA measurements of properties like spin
alignment can tell us about the surrounding gas (or lack thereof).  It
may be possible, from the gravitational waves alone, to predict the
nature of an electromagnetic counterpart or to make statements about a
binary's environment in case a counterpart is missed.  We plan to
study this issue in more detail in the future.

Lacking from our Fisher-matrix-based study is an investigation of how
the spin-alignment priors of the hot, cold, and dry scenarios may
affect parameter estimation.  In a more thorough Bayesian analysis,
the range of spin alignments for each model can be included as priors.
Bayesian model selection can then be used to identify the model that
best describes the data.  For cold gas mergers, the tight priors on
the spin-orbit alignment would translate into improvements in the
parameter estimation over what we have found here.  Thus, the spin-alignment priors can help put back some of parameter recovery accuracy
that is taken away by the suppression of the spin precession.  We are
currently investigating this issue.

An obvious avenue for LISA parameter estimation studies is the
improvement of the waveform model.  While the inspiral is nearly
complete, the study of the merger is just beginning.  In the next few
years, complete inspiral waveforms from codes like this one will be
joined to effective-one-body waveforms for the late inspiral,
fits to numerical models of the merger, and perturbative ringdowns to
give a complete LISA waveform for analysis purposes.  While such
studies have begun \cite{mtbk10,mlbt11}, they do not yet include spins.
Including the merger and ringdown is critical; not only do they
provide a great deal of SNR and parameter information, but they also
provide a physical tapering of the waveform.  The results of this
paper unfortunately do not always quantitatively match those of Paper
I and \cite{kjs09} when appropriate, primarily because the earlier
studies use the stationary phase approximation while we taper the
signal and apply an FFT.  In essence, despite the use of the MECO, we
are applying an earlier cutoff and losing some information about our
parameters.  When the complete signal is used, choices of cutoff will
become irrelevant, and different results should agree more readily.
Of course, for this paper, exact error estimates are not the end goal;
instead, we have aimed only to show general behavior and trends which
are independent of any shift in the baseline error values.

Finally, the Fisher-matrix formalism itself must be checked to make
sure it is correct with complicated waveforms and large numbers of
parameters.  We plan to carry out a comparison between this code's
results and those obtained by exploring the full posterior probability
using Markov Chain Monte Carlo techniques
\cite{cm01,cc05,cp06,cp07}.  Early results have shown that the Fisher
matrix is indeed still valid in most regimes --- a relief given that Markov
Chain Monte Carlo techniques cannot as easily survey a plethora of sky locations and
orientations --- but the full parameter space has not been explored.
It is clear, however, that care must be taken with all three areas,
astrophysics, general relativity, and statistical analysis, before a
final picture of LISA science capabilities can be established.

\acknowledgments We thank Cole Miller for suggesting the problem to
us.  We also thank Samaya Nissanke, Sean McWilliams, and Tyson
Littenberg for useful discussions.  R.N.L. was supported by an
appointment to the NASA Postdoctoral Program at the Goddard Space
Flight Center, administered by Oak Ridge Associated Universities
through a contract with NASA.  This work is supported at MIT by NASA
Grant NNX08AL42G and NSF Grant PHY--0449884.  S.A.H. in addition
gratefully acknowledges the support of the Adam J.\ Burgasser Chair in
Astrophysics in completing this analysis. N.J.C. was supported by NASA
Grant NNX10AH15G.

\vfill

\bibliographystyle{apsrev}
\bibliography{alignedspins}

\newcommand{\noopsort}[1]{} \newcommand{\printfirst}[2]{#1}
  \newcommand{\singleletter}[1]{#1} \newcommand{\switchargs}[2]{#2#1}
\begin{thebibliography}{51}
\expandafter\ifx\csname natexlab\endcsname\relax\def\natexlab#1{#1}\fi
\expandafter\ifx\csname bibnamefont\endcsname\relax
  \def\bibnamefont#1{#1}\fi
\expandafter\ifx\csname bibfnamefont\endcsname\relax
  \def\bibfnamefont#1{#1}\fi
\expandafter\ifx\csname citenamefont\endcsname\relax
  \def\citenamefont#1{#1}\fi
\expandafter\ifx\csname url\endcsname\relax
  \def\url#1{\texttt{#1}}\fi
\expandafter\ifx\csname urlprefix\endcsname\relax\def\urlprefix{URL }\fi
\providecommand{\bibinfo}[2]{#2}
\providecommand{\eprint}[2][]{\url{#2}}

\bibitem[{\citenamefont{{Baker} et~al.}(2007)\citenamefont{{Baker},
  {McWilliams}, {van Meter}, {Centrella}, {Choi}, {Kelly}, and
  {Koppitz}}}]{bmvcckk07}
\bibinfo{author}{\bibfnamefont{J.~G.} \bibnamefont{{Baker}}},
  \bibinfo{author}{\bibfnamefont{S.~T.} \bibnamefont{{McWilliams}}},
  \bibinfo{author}{\bibfnamefont{J.~R.} \bibnamefont{{van Meter}}},
  \bibinfo{author}{\bibfnamefont{J.}~\bibnamefont{{Centrella}}},
  \bibinfo{author}{\bibfnamefont{D.-I.} \bibnamefont{{Choi}}},
  \bibinfo{author}{\bibfnamefont{B.~J.} \bibnamefont{{Kelly}}},
  \bibnamefont{and}
  \bibinfo{author}{\bibfnamefont{M.}~\bibnamefont{{Koppitz}}},
  \bibinfo{journal}{\prd} \textbf{\bibinfo{volume}{75}},
  \bibinfo{pages}{124024} (\bibinfo{year}{2007}).

\bibitem[{\citenamefont{{Sesana} et~al.}(2007)\citenamefont{{Sesana},
  {Volonteri}, and {Haardt}}}]{svh07}
\bibinfo{author}{\bibfnamefont{A.}~\bibnamefont{{Sesana}}},
  \bibinfo{author}{\bibfnamefont{M.}~\bibnamefont{{Volonteri}}},
  \bibnamefont{and} \bibinfo{author}{\bibfnamefont{F.}~\bibnamefont{{Haardt}}},
  \bibinfo{journal}{Mon.\ Not.\ R.\ Astron.\ Soc.}
  \textbf{\bibinfo{volume}{377}}, \bibinfo{pages}{1711} (\bibinfo{year}{2007}).

\bibitem[{\citenamefont{{Cutler}}(1998)}]{c98}
\bibinfo{author}{\bibfnamefont{C.}~\bibnamefont{{Cutler}}},
  \bibinfo{journal}{\prd} \textbf{\bibinfo{volume}{57}}, \bibinfo{pages}{7089}
  (\bibinfo{year}{1998}).

\bibitem[{\citenamefont{{Hughes}}(2002)}]{h02}
\bibinfo{author}{\bibfnamefont{S.~A.} \bibnamefont{{Hughes}}},
  \bibinfo{journal}{Mon.\ Not.\ R.\ Astron.\ Soc.}
  \textbf{\bibinfo{volume}{331}}, \bibinfo{pages}{805} (\bibinfo{year}{2002}).

\bibitem[{\citenamefont{{Vecchio}}(2004)}]{v04}
\bibinfo{author}{\bibfnamefont{A.}~\bibnamefont{{Vecchio}}},
  \bibinfo{journal}{\prd} \textbf{\bibinfo{volume}{70}},
  \bibinfo{pages}{042001} (\bibinfo{year}{2004}).

\bibitem[{\citenamefont{{Berti} et~al.}(2005)\citenamefont{{Berti}, {Buonanno},
  and {Will}}}]{bbw05}
\bibinfo{author}{\bibfnamefont{E.}~\bibnamefont{{Berti}}},
  \bibinfo{author}{\bibfnamefont{A.}~\bibnamefont{{Buonanno}}},
  \bibnamefont{and} \bibinfo{author}{\bibfnamefont{C.~M.}
  \bibnamefont{{Will}}}, \bibinfo{journal}{\prd} \textbf{\bibinfo{volume}{71}},
  \bibinfo{pages}{084025} (\bibinfo{year}{2005}).

\bibitem[{\citenamefont{{Holz} and {Hughes}}(2005)}]{hh05}
\bibinfo{author}{\bibfnamefont{D.~E.} \bibnamefont{{Holz}}} \bibnamefont{and}
  \bibinfo{author}{\bibfnamefont{S.~A.} \bibnamefont{{Hughes}}},
  \bibinfo{journal}{\apj} \textbf{\bibinfo{volume}{629}}, \bibinfo{pages}{15}
  (\bibinfo{year}{2005}).

\bibitem[{\citenamefont{{Berti} et~al.}(2006)\citenamefont{{Berti}, {Cardoso},
  and {Will}}}]{bcw06}
\bibinfo{author}{\bibfnamefont{E.}~\bibnamefont{{Berti}}},
  \bibinfo{author}{\bibfnamefont{V.}~\bibnamefont{{Cardoso}}},
  \bibnamefont{and} \bibinfo{author}{\bibfnamefont{C.~M.}
  \bibnamefont{{Will}}}, \bibinfo{journal}{\prd} \textbf{\bibinfo{volume}{73}},
  \bibinfo{pages}{064030} (\bibinfo{year}{2006}).

\bibitem[{\citenamefont{{Lang} and {Hughes}}()}]{lh06}
\bibinfo{author}{\bibfnamefont{R.~N.} \bibnamefont{{Lang}}} \bibnamefont{and}
  \bibinfo{author}{\bibfnamefont{S.~A.} \bibnamefont{{Hughes}}},
  \bibinfo{note}{\prd {\bf 74}, 122001 (2006); and , {\bf 75}, 089902(E)
  (2007); and , {\bf 77}, 109901(E) (2008)}.

\bibitem[{\citenamefont{{Arun} et~al.}(2007{\natexlab{a}})\citenamefont{{Arun},
  {Iyer}, {Sathyaprakash}, and {Sinha}}}]{aiss07}
\bibinfo{author}{\bibfnamefont{K.~G.} \bibnamefont{{Arun}}},
  \bibinfo{author}{\bibfnamefont{B.~R.} \bibnamefont{{Iyer}}},
  \bibinfo{author}{\bibfnamefont{B.~S.} \bibnamefont{{Sathyaprakash}}},
  \bibnamefont{and} \bibinfo{author}{\bibfnamefont{S.}~\bibnamefont{{Sinha}}},
  \bibinfo{journal}{\prd} \textbf{\bibinfo{volume}{75}},
  \bibinfo{pages}{124002} (\bibinfo{year}{2007}{\natexlab{a}}).

\bibitem[{\citenamefont{{Arun} et~al.}(2007{\natexlab{b}})\citenamefont{{Arun},
  {Iyer}, {Sathyaprakash}, {Sinha}, and {van den Broeck}}}]{aissv07}
\bibinfo{author}{\bibfnamefont{K.~G.} \bibnamefont{{Arun}}},
  \bibinfo{author}{\bibfnamefont{B.~R.} \bibnamefont{{Iyer}}},
  \bibinfo{author}{\bibfnamefont{B.~S.} \bibnamefont{{Sathyaprakash}}},
  \bibinfo{author}{\bibfnamefont{S.}~\bibnamefont{{Sinha}}}, \bibnamefont{and}
  \bibinfo{author}{\bibfnamefont{C.}~\bibnamefont{{van den Broeck}}},
  \bibinfo{journal}{\prd} \textbf{\bibinfo{volume}{76}},
  \bibinfo{pages}{104016} (\bibinfo{year}{2007}{\natexlab{b}}).

\bibitem[{\citenamefont{{Trias} and {Sintes}}(2008)}]{ts08}
\bibinfo{author}{\bibfnamefont{M.}~\bibnamefont{{Trias}}} \bibnamefont{and}
  \bibinfo{author}{\bibfnamefont{A.~M.} \bibnamefont{{Sintes}}},
  \bibinfo{journal}{\prd} \textbf{\bibinfo{volume}{77}},
  \bibinfo{pages}{024030} (\bibinfo{year}{2008}).

\bibitem[{\citenamefont{{Lang} and {Hughes}}(2008)}]{lh08}
\bibinfo{author}{\bibfnamefont{R.~N.} \bibnamefont{{Lang}}} \bibnamefont{and}
  \bibinfo{author}{\bibfnamefont{S.~A.} \bibnamefont{{Hughes}}},
  \bibinfo{journal}{\apj} \textbf{\bibinfo{volume}{677}}, \bibinfo{pages}{1184}
  (\bibinfo{year}{2008}).

\bibitem[{\citenamefont{{Porter} and {Cornish}}(2008)}]{pc08}
\bibinfo{author}{\bibfnamefont{E.~K.} \bibnamefont{{Porter}}} \bibnamefont{and}
  \bibinfo{author}{\bibfnamefont{N.~J.} \bibnamefont{{Cornish}}},
  \bibinfo{journal}{\prd} \textbf{\bibinfo{volume}{78}},
  \bibinfo{pages}{064005} (\bibinfo{year}{2008}).

\bibitem[{\citenamefont{{Lang} and {Hughes}}(2009)}]{lh09}
\bibinfo{author}{\bibfnamefont{R.~N.} \bibnamefont{{Lang}}} \bibnamefont{and}
  \bibinfo{author}{\bibfnamefont{S.~A.} \bibnamefont{{Hughes}}},
  \bibinfo{journal}{Classical Quantum Gravity} \textbf{\bibinfo{volume}{26}},
  \bibinfo{pages}{094035} (\bibinfo{year}{2009}).

\bibitem[{\citenamefont{{Klein} et~al.}(2009)\citenamefont{{Klein}, {Jetzer},
  and {Sereno}}}]{kjs09}
\bibinfo{author}{\bibfnamefont{A.}~\bibnamefont{{Klein}}},
  \bibinfo{author}{\bibfnamefont{P.}~\bibnamefont{{Jetzer}}}, \bibnamefont{and}
  \bibinfo{author}{\bibfnamefont{M.}~\bibnamefont{{Sereno}}},
  \bibinfo{journal}{\prd} \textbf{\bibinfo{volume}{80}},
  \bibinfo{pages}{064027} (\bibinfo{year}{2009}).

\bibitem[{\citenamefont{{McWilliams} et~al.}(2010)\citenamefont{{McWilliams},
  {Thorpe}, {Baker}, and {Kelly}}}]{mtbk10}
\bibinfo{author}{\bibfnamefont{S.~T.} \bibnamefont{{McWilliams}}},
  \bibinfo{author}{\bibfnamefont{J.~I.} \bibnamefont{{Thorpe}}},
  \bibinfo{author}{\bibfnamefont{J.~G.} \bibnamefont{{Baker}}},
  \bibnamefont{and} \bibinfo{author}{\bibfnamefont{B.~J.}
  \bibnamefont{{Kelly}}}, \bibinfo{journal}{\prd}
  \textbf{\bibinfo{volume}{81}}, \bibinfo{pages}{064014}
  (\bibinfo{year}{2010}).

\bibitem[{\citenamefont{Key and Cornish}(2011)}]{kc11}
\bibinfo{author}{\bibfnamefont{J.~S.} \bibnamefont{Key}} \bibnamefont{and}
  \bibinfo{author}{\bibfnamefont{N.~J.} \bibnamefont{Cornish}},
  \bibinfo{journal}{Phys. Rev. D} \textbf{\bibinfo{volume}{83}},
  \bibinfo{pages}{083001} (\bibinfo{year}{2011}).

\bibitem[{\citenamefont{{McWilliams} et~al.}()\citenamefont{{McWilliams},
  {Lang}, {Baker}, and {Thorpe}}}]{mlbt11}
\bibinfo{author}{\bibfnamefont{S.~T.} \bibnamefont{{McWilliams}}},
  \bibinfo{author}{\bibfnamefont{R.~N.} \bibnamefont{{Lang}}},
  \bibinfo{author}{\bibfnamefont{J.~G.} \bibnamefont{{Baker}}},
  \bibnamefont{and} \bibinfo{author}{\bibfnamefont{J.~I.}
  \bibnamefont{{Thorpe}}}, \eprint{arXiv:1104.5650}.

\bibitem[{\citenamefont{{Finn}}(1992)}]{f92}
\bibinfo{author}{\bibfnamefont{L.~S.} \bibnamefont{{Finn}}},
  \bibinfo{journal}{\prd} \textbf{\bibinfo{volume}{46}}, \bibinfo{pages}{5236}
  (\bibinfo{year}{1992}).

\bibitem[{\citenamefont{{Cutler} and {Flanagan}}(1994)}]{cf94}
\bibinfo{author}{\bibfnamefont{C.}~\bibnamefont{{Cutler}}} \bibnamefont{and}
  \bibinfo{author}{\bibfnamefont{{\'E}.~{\'E}.} \bibnamefont{{Flanagan}}},
  \bibinfo{journal}{\prd} \textbf{\bibinfo{volume}{49}}, \bibinfo{pages}{2658}
  (\bibinfo{year}{1994}).

\bibitem[{\citenamefont{{K.~A.\ Arnaud, S.\ Babak, J.~G.\ Baker, M.~J.
  Benacquista, N.~J.\ Cornish, C.\ Cutler, S.~L.\ Larson, B.~S. Sathyaprakash,
  M.\ Vallisneri, A.\ Vecchio {\it et al.}}}(2006)}]{a06}
\bibinfo{author}{\bibnamefont{{K.~A.\ Arnaud, S.\ Babak, J.~G.\ Baker, M.~J.
  Benacquista, N.~J.\ Cornish, C.\ Cutler, S.~L.\ Larson, B.~S. Sathyaprakash,
  M.\ Vallisneri, A.\ Vecchio {\it et al.}}}}, in
  \emph{\bibinfo{booktitle}{{Laser Interferometer Space Antenna: 6th
  International LISA Symposium, {\rm American Institute of Physics Conference
  Series Vol.\ 873}}}}, edited by
  \bibinfo{editor}{\bibnamefont{{S.~M.~Merkowitz and J.~C.~Livas}}}
  (\bibinfo{year}{2006}), pp. \bibinfo{pages}{619--624}.

\bibitem[{\citenamefont{{Armitage} and {Natarajan}}(2002)}]{an02}
\bibinfo{author}{\bibfnamefont{P.~J.} \bibnamefont{{Armitage}}}
  \bibnamefont{and}
  \bibinfo{author}{\bibfnamefont{P.}~\bibnamefont{{Natarajan}}},
  \bibinfo{journal}{Astrophys.\ J.\ Lett.} \textbf{\bibinfo{volume}{567}},
  \bibinfo{pages}{L9} (\bibinfo{year}{2002}).

\bibitem[{\citenamefont{{Bode} and {Phinney}}(2009)}]{bp09}
\bibinfo{author}{\bibfnamefont{J.~N.} \bibnamefont{{Bode}}} \bibnamefont{and}
  \bibinfo{author}{\bibfnamefont{E.}~\bibnamefont{{Phinney}}},
  \bibinfo{journal}{Bull.\ Am.\ Astron.\ Soc.} \textbf{\bibinfo{volume}{41}},
  \bibinfo{pages}{341} (\bibinfo{year}{2009}).

\bibitem[{\citenamefont{{O'Neill} et~al.}(2009)\citenamefont{{O'Neill},
  {Miller}, {Bogdanovi{\'c}}, {Reynolds}, and {Schnittman}}}]{ombrs09}
\bibinfo{author}{\bibfnamefont{S.~M.} \bibnamefont{{O'Neill}}},
  \bibinfo{author}{\bibfnamefont{M.~C.} \bibnamefont{{Miller}}},
  \bibinfo{author}{\bibfnamefont{T.}~\bibnamefont{{Bogdanovi{\'c}}}},
  \bibinfo{author}{\bibfnamefont{C.~S.} \bibnamefont{{Reynolds}}},
  \bibnamefont{and} \bibinfo{author}{\bibfnamefont{J.~D.}
  \bibnamefont{{Schnittman}}}, \bibinfo{journal}{\apj}
  \textbf{\bibinfo{volume}{700}}, \bibinfo{pages}{859} (\bibinfo{year}{2009}).

\bibitem[{\citenamefont{{Milosavljevi{\'c}} and {Phinney}}(2005)}]{mp05}
\bibinfo{author}{\bibfnamefont{M.}~\bibnamefont{{Milosavljevi{\'c}}}}
  \bibnamefont{and} \bibinfo{author}{\bibfnamefont{E.~S.}
  \bibnamefont{{Phinney}}}, \bibinfo{journal}{Astrophys.\ J.\ Lett.}
  \textbf{\bibinfo{volume}{622}}, \bibinfo{pages}{L93} (\bibinfo{year}{2005}).

\bibitem[{\citenamefont{{Stone} and {Loeb}}(2011)}]{sl11}
\bibinfo{author}{\bibfnamefont{N.}~\bibnamefont{{Stone}}} \bibnamefont{and}
  \bibinfo{author}{\bibfnamefont{A.}~\bibnamefont{{Loeb}}},
  \bibinfo{journal}{Mon.\ Not.\ R.\ Astron.\ Soc.}
  \textbf{\bibinfo{volume}{412}}, \bibinfo{pages}{75} (\bibinfo{year}{2011}).

\bibitem[{\citenamefont{{Wegg} and {Bode}}()}]{wb10}
\bibinfo{author}{\bibfnamefont{C.}~\bibnamefont{{Wegg}}} \bibnamefont{and}
  \bibinfo{author}{\bibfnamefont{J.~N.} \bibnamefont{{Bode}}},
  \eprint{arXiv:1011.5874}.

\bibitem[{\citenamefont{{Tyson}}({SPIE, Bellingham, WA, 2002})}]{t02}
\bibinfo{author}{\bibfnamefont{J.~A.} \bibnamefont{{Tyson}}}, in
  \emph{\bibinfo{booktitle}{Survey and Other Telescope Technologies and
  Discoveries}}, edited by \bibinfo{editor}{\bibnamefont{{J.~A.\ Tyson and S.\
  Wolff, SPIE Conference Series, Vol.\ 4836}}} (\bibinfo{year}{{SPIE,
  Bellingham, WA, 2002}}), pp. \bibinfo{pages}{10--20}.

\bibitem[{\citenamefont{{Apostolatos} et~al.}(1994)\citenamefont{{Apostolatos},
  {Cutler}, {Sussman}, and {Thorne}}}]{acst94}
\bibinfo{author}{\bibfnamefont{T.~A.} \bibnamefont{{Apostolatos}}},
  \bibinfo{author}{\bibfnamefont{C.}~\bibnamefont{{Cutler}}},
  \bibinfo{author}{\bibfnamefont{G.~J.} \bibnamefont{{Sussman}}},
  \bibnamefont{and} \bibinfo{author}{\bibfnamefont{K.~S.}
  \bibnamefont{{Thorne}}}, \bibinfo{journal}{\prd}
  \textbf{\bibinfo{volume}{49}}, \bibinfo{pages}{6274} (\bibinfo{year}{1994}).

\bibitem[{\citenamefont{{Kidder}}(1995)}]{k95}
\bibinfo{author}{\bibfnamefont{L.~E.} \bibnamefont{{Kidder}}},
  \bibinfo{journal}{\prd} \textbf{\bibinfo{volume}{52}}, \bibinfo{pages}{821}
  (\bibinfo{year}{1995}).

\bibitem[{\citenamefont{{Bogdanovi{\'c}}
  et~al.}(2007)\citenamefont{{Bogdanovi{\'c}}, {Reynolds}, and
  {Miller}}}]{brm07}
\bibinfo{author}{\bibfnamefont{T.}~\bibnamefont{{Bogdanovi{\'c}}}},
  \bibinfo{author}{\bibfnamefont{C.~S.} \bibnamefont{{Reynolds}}},
  \bibnamefont{and} \bibinfo{author}{\bibfnamefont{M.~C.}
  \bibnamefont{{Miller}}}, \bibinfo{journal}{Astrophys.\ J.\ Lett.}
  \textbf{\bibinfo{volume}{661}}, \bibinfo{pages}{L147} (\bibinfo{year}{2007}).

\bibitem[{\citenamefont{{Dotti} et~al.}(2010)\citenamefont{{Dotti},
  {Volonteri}, {Perego}, {Colpi}, {Ruszkowski}, and {Haardt}}}]{d10}
\bibinfo{author}{\bibfnamefont{M.}~\bibnamefont{{Dotti}}},
  \bibinfo{author}{\bibfnamefont{M.}~\bibnamefont{{Volonteri}}},
  \bibinfo{author}{\bibfnamefont{A.}~\bibnamefont{{Perego}}},
  \bibinfo{author}{\bibfnamefont{M.}~\bibnamefont{{Colpi}}},
  \bibinfo{author}{\bibfnamefont{M.}~\bibnamefont{{Ruszkowski}}},
  \bibnamefont{and} \bibinfo{author}{\bibfnamefont{F.}~\bibnamefont{{Haardt}}},
  \bibinfo{journal}{Mon.\ Not.\ R.\ Astron.\ Soc.}
  \textbf{\bibinfo{volume}{402}}, \bibinfo{pages}{682} (\bibinfo{year}{2010}).

\bibitem[{\citenamefont{{K.~G.\ Arun, S.\ Babak, E.\ Berti, N.\ Cornish, C.\
  Cutler, J.\ Gair, S.~A.\ Hughes, B.~R.\ Iyer, R.~N.\ Lang, I.\ Mandel {\it et
  al.}}}(2009)}]{a09}
\bibinfo{author}{\bibnamefont{{K.~G.\ Arun, S.\ Babak, E.\ Berti, N.\ Cornish,
  C.\ Cutler, J.\ Gair, S.~A.\ Hughes, B.~R.\ Iyer, R.~N.\ Lang, I.\ Mandel
  {\it et al.}}}}, \bibinfo{journal}{Classical Quantum Gravity}
  \textbf{\bibinfo{volume}{26}}, \bibinfo{pages}{094027}
  (\bibinfo{year}{2009}).

\bibitem[{\citenamefont{{Peters}}(1964)}]{p64}
\bibinfo{author}{\bibfnamefont{P.~C.} \bibnamefont{{Peters}}},
  \bibinfo{journal}{Phys.\ Rev.} \textbf{\bibinfo{volume}{136}},
  \bibinfo{pages}{B1224} (\bibinfo{year}{1964}).

\bibitem[{\citenamefont{{Armitage} and {Natarajan}}(2005)}]{an05}
\bibinfo{author}{\bibfnamefont{P.~J.} \bibnamefont{{Armitage}}}
  \bibnamefont{and}
  \bibinfo{author}{\bibfnamefont{P.}~\bibnamefont{{Natarajan}}},
  \bibinfo{journal}{\apj} \textbf{\bibinfo{volume}{634}}, \bibinfo{pages}{921}
  (\bibinfo{year}{2005}).

\bibitem[{\citenamefont{{Cuadra} et~al.}(2009)\citenamefont{{Cuadra},
  {Armitage}, {Alexander}, and {Begelman}}}]{caab09}
\bibinfo{author}{\bibfnamefont{J.}~\bibnamefont{{Cuadra}}},
  \bibinfo{author}{\bibfnamefont{P.~J.} \bibnamefont{{Armitage}}},
  \bibinfo{author}{\bibfnamefont{R.~D.} \bibnamefont{{Alexander}}},
  \bibnamefont{and} \bibinfo{author}{\bibfnamefont{M.~C.}
  \bibnamefont{{Begelman}}}, \bibinfo{journal}{Mon.\ Not.\ R.\ Astron.\ Soc.}
  \textbf{\bibinfo{volume}{393}}, \bibinfo{pages}{1423} (\bibinfo{year}{2009}).

\bibitem[{\citenamefont{{Sesana}}(2010)}]{s10}
\bibinfo{author}{\bibfnamefont{A.}~\bibnamefont{{Sesana}}},
  \bibinfo{journal}{\apj} \textbf{\bibinfo{volume}{719}}, \bibinfo{pages}{851}
  (\bibinfo{year}{2010}).

\bibitem[{\citenamefont{{Buonanno} et~al.}(2003)\citenamefont{{Buonanno},
  {Chen}, and {Vallisneri}}}]{bcv03}
\bibinfo{author}{\bibfnamefont{A.}~\bibnamefont{{Buonanno}}},
  \bibinfo{author}{\bibfnamefont{Y.}~\bibnamefont{{Chen}}}, \bibnamefont{and}
  \bibinfo{author}{\bibfnamefont{M.}~\bibnamefont{{Vallisneri}}},
  \bibinfo{journal}{\prd} \textbf{\bibinfo{volume}{67}},
  \bibinfo{pages}{104025} (\bibinfo{year}{2003}).

\bibitem[{\citenamefont{{Boyle} et~al.}(2007)\citenamefont{{Boyle}, {Brown},
  {Kidder}, {Mrou{\'e}}, {Pfeiffer}, {Scheel}, {Cook}, and {Teukolsky}}}]{b07}
\bibinfo{author}{\bibfnamefont{M.}~\bibnamefont{{Boyle}}},
  \bibinfo{author}{\bibfnamefont{D.~A.} \bibnamefont{{Brown}}},
  \bibinfo{author}{\bibfnamefont{L.~E.} \bibnamefont{{Kidder}}},
  \bibinfo{author}{\bibfnamefont{A.~H.} \bibnamefont{{Mrou{\'e}}}},
  \bibinfo{author}{\bibfnamefont{H.~P.} \bibnamefont{{Pfeiffer}}},
  \bibinfo{author}{\bibfnamefont{M.~A.} \bibnamefont{{Scheel}}},
  \bibinfo{author}{\bibfnamefont{G.~B.} \bibnamefont{{Cook}}},
  \bibnamefont{and} \bibinfo{author}{\bibfnamefont{S.~A.}
  \bibnamefont{{Teukolsky}}}, \bibinfo{journal}{\prd}
  \textbf{\bibinfo{volume}{76}}, \bibinfo{pages}{124038}
  (\bibinfo{year}{2007}).

\bibitem[{\citenamefont{{O'Sullivan} and {Hughes}}()}]{osh11}
\bibinfo{author}{\bibfnamefont{S.~G.} \bibnamefont{{O'Sullivan}}}
  \bibnamefont{and} \bibinfo{author}{\bibfnamefont{S.~A.}
  \bibnamefont{{Hughes}}}, \bibinfo{note}{\prd (to be published)}.

\bibitem[{\citenamefont{{Rubbo} et~al.}(2004)\citenamefont{{Rubbo}, {Cornish},
  and {Poujade}}}]{rcp04}
\bibinfo{author}{\bibfnamefont{L.~J.} \bibnamefont{{Rubbo}}},
  \bibinfo{author}{\bibfnamefont{N.~J.} \bibnamefont{{Cornish}}},
  \bibnamefont{and}
  \bibinfo{author}{\bibfnamefont{O.}~\bibnamefont{{Poujade}}},
  \bibinfo{journal}{\prd} \textbf{\bibinfo{volume}{69}},
  \bibinfo{pages}{082003} (\bibinfo{year}{2004}).

\bibitem[{\citenamefont{{Vallisneri}}(2005)}]{v05}
\bibinfo{author}{\bibfnamefont{M.}~\bibnamefont{{Vallisneri}}},
  \bibinfo{journal}{\prd} \textbf{\bibinfo{volume}{71}},
  \bibinfo{pages}{022001} (\bibinfo{year}{2005}).

\bibitem[{\citenamefont{{Petiteau} et~al.}(2008)\citenamefont{{Petiteau},
  {Auger}, {Halloin}, {Jeannin}, {Plagnol}, {Pireaux}, {Regimbau}, and
  {Vinet}}}]{pahjpprv08}
\bibinfo{author}{\bibfnamefont{A.}~\bibnamefont{{Petiteau}}},
  \bibinfo{author}{\bibfnamefont{G.}~\bibnamefont{{Auger}}},
  \bibinfo{author}{\bibfnamefont{H.}~\bibnamefont{{Halloin}}},
  \bibinfo{author}{\bibfnamefont{O.}~\bibnamefont{{Jeannin}}},
  \bibinfo{author}{\bibfnamefont{E.}~\bibnamefont{{Plagnol}}},
  \bibinfo{author}{\bibfnamefont{S.}~\bibnamefont{{Pireaux}}},
  \bibinfo{author}{\bibfnamefont{T.}~\bibnamefont{{Regimbau}}},
  \bibnamefont{and} \bibinfo{author}{\bibfnamefont{J.}~\bibnamefont{{Vinet}}},
  \bibinfo{journal}{\prd} \textbf{\bibinfo{volume}{77}},
  \bibinfo{pages}{023002} (\bibinfo{year}{2008}).

\bibitem[{\citenamefont{{Armstrong} et~al.}(1999)\citenamefont{{Armstrong},
  {Estabrook}, and {Tinto}}}]{aet99}
\bibinfo{author}{\bibfnamefont{J.~W.} \bibnamefont{{Armstrong}}},
  \bibinfo{author}{\bibfnamefont{F.~B.} \bibnamefont{{Estabrook}}},
  \bibnamefont{and} \bibinfo{author}{\bibfnamefont{M.}~\bibnamefont{{Tinto}}},
  \bibinfo{journal}{\apj} \textbf{\bibinfo{volume}{527}}, \bibinfo{pages}{814}
  (\bibinfo{year}{1999}).

\bibitem[{\citenamefont{{Crowder} and {Cornish}}(2007)}]{cc07}
\bibinfo{author}{\bibfnamefont{J.}~\bibnamefont{{Crowder}}} \bibnamefont{and}
  \bibinfo{author}{\bibfnamefont{N.~J.} \bibnamefont{{Cornish}}},
  \bibinfo{journal}{\prd} \textbf{\bibinfo{volume}{75}},
  \bibinfo{pages}{043008} (\bibinfo{year}{2007}).

\bibitem[{\citenamefont{{Blanchet}}(2006)}]{b06}
\bibinfo{author}{\bibfnamefont{L.}~\bibnamefont{{Blanchet}}},
  \bibinfo{journal}{Living Rev.\ Relativity} \textbf{\bibinfo{volume}{9}},
  \bibinfo{pages}{4} (\bibinfo{year}{2006}),
  \bibinfo{note}{{http://relativity.livingreviews.org/Articles/lrr-2006-4/}}.

\bibitem[{\citenamefont{{Christensen} and {Meyer}}(2001)}]{cm01}
\bibinfo{author}{\bibfnamefont{N.}~\bibnamefont{{Christensen}}}
  \bibnamefont{and} \bibinfo{author}{\bibfnamefont{R.}~\bibnamefont{{Meyer}}},
  \bibinfo{journal}{\prd} \textbf{\bibinfo{volume}{64}},
  \bibinfo{pages}{022001} (\bibinfo{year}{2001}).

\bibitem[{\citenamefont{{Cornish} and {Crowder}}(2005)}]{cc05}
\bibinfo{author}{\bibfnamefont{N.~J.} \bibnamefont{{Cornish}}}
  \bibnamefont{and}
  \bibinfo{author}{\bibfnamefont{J.}~\bibnamefont{{Crowder}}},
  \bibinfo{journal}{\prd} \textbf{\bibinfo{volume}{72}},
  \bibinfo{pages}{043005} (\bibinfo{year}{2005}).

\bibitem[{\citenamefont{{Cornish} and {Porter}}(2006)}]{cp06}
\bibinfo{author}{\bibfnamefont{N.~J.} \bibnamefont{{Cornish}}}
  \bibnamefont{and} \bibinfo{author}{\bibfnamefont{E.~K.}
  \bibnamefont{{Porter}}}, \bibinfo{journal}{Classical Quantum Gravity}
  \textbf{\bibinfo{volume}{23}}, \bibinfo{pages}{S761} (\bibinfo{year}{2006}).

\bibitem[{\citenamefont{{Cornish} and {Porter}}(2007)}]{cp07}
\bibinfo{author}{\bibfnamefont{N.~J.} \bibnamefont{{Cornish}}}
  \bibnamefont{and} \bibinfo{author}{\bibfnamefont{E.~K.}
  \bibnamefont{{Porter}}}, \bibinfo{journal}{\prd}
  \textbf{\bibinfo{volume}{75}}, \bibinfo{pages}{021301}
  (\bibinfo{year}{2007}).

\end{thebibliography}

\end{document}